\documentclass{aa}  
\usepackage[switch]{lineno}
\usepackage{amsmath}
\usepackage{booktabs}
\usepackage{threeparttable}
\usepackage{multirow}
\usepackage{tabularx}
\usepackage{threeparttable}
\usepackage{graphicx}
\usepackage{txfonts}
\usepackage[]{hyperref}
\hypersetup{unicode=true, colorlinks=true, linkcolor=[rgb]{0.0, 0.0, 1}, citecolor=[rgb]{0.0, 0.0, 1}, filecolor=[rgb]{0.0, 0.0, 1}, urlcolor=[rgb]{0.0, 0.0, 1}}
\usepackage{natbib}
\usepackage{scrhack} 
\begin{document} 

\title{Cross-spectral Analysis of the Type-C Quasi-periodic Oscillation Shoulder Component in GX 339$-$4}


     \author{Haifan Zhu
          \inst{1}\fnmsep\thanks{zhu@astro.rug.nl}
          \and
          Mariano M\'endez\inst{1}\fnmsep\thanks{mariano@astro.rug.nl}
          \and
          Pengcheng Yang\inst{1}
          \and
          Pei Jin\inst{1}\and
          Candela Bellavita\inst{1,2,3}
          \and
          Federico Garc\'ia\inst{2,3}
          \and 
           Wei Wang\inst{4}
          \and 
          Diego Altamirano\inst{5}
          \and
          Liang Zhang \inst{6}
          \and
          Chenxu Gao\inst{7}
          \and
          Xiao Chen\inst{8}}

   \institute{Kapteyn Astronomical Institute, University of Groningen, P.O. BOX 800, 9700 AV Groningen, The Netherlands
   \and 
   Instituto Argentino de Radioastronom\'ia (CCT La Plata, CONICET; CICPBA; UNLP), C.C.5, 1894 Villa Elisa, Argentina
   \and Facultad de Ciencias Astron\'omicas y Geof\'isicas, Universidad Nacional de La Plata, 1900 La Plata, Argentina
   \and
   Department of Astronomy, School of Physics and Technology, Wuhan University, Wuhan 430072, China
   \and
   School of Physics and Astronomy, University of Southampton, Southampton, Hampshire SO17 1BJ, UK
   \and 
   Key Laboratory for Particle Astrophysics, Institute of High Energy Physics, Chinese Academy of Sciences, 19B Yuquan Road,
Beijing 100049, China
\and             
Shanghai Astronomical Observatory, Chinese Academy of Sciences, Nandan Road, Shanghai 200030, China   
\and
Department of Physics, College of Science, Southern University of Science and Technology, Shenzhen 518055, China
         }

   \date{}

\date{Received XXX; accepted YYY}

\abstract
{
We revisit \textit{Rossi X-ray Timing Explorer} (\textit{RXTE}) observations of GX~339$-$4 during the rising phase of its 2006/2007 outburst and apply a joint power-density-spectrum (PDS)--cross-spectrum (CS) decomposition to the type-C quasi-periodic oscillation (QPO) region.
Within this framework, the QPO region is described by a narrow QPO
fundamental and a neighboring high-frequency shoulder, whose amplitudes
and phase lags can be measured separately. 
The shoulder is first detected at MJD~54142.04, mainly through the imaginary part of the CS and a narrow local structure in the phase-lag spectrum, before becoming a resolved high-frequency shoulder in the PDS.
It follows the QPO frequency evolution on the high-frequency side, with $R_{\nu}=\nu_{\rm sh}/\nu_{\rm QPO}\simeq1.04$--$1.18$.
The QPO lag remains small, typically below $\sim0.17$~rad, whereas the shoulder carries a larger hard lag of $\sim0.5$--$0.8$~rad. 
 Energy-resolved fits show
the same separation: the QPO lag is close to zero or only weakly positive across most of the energy band, while the shoulder lag
is systematically larger and generally increases with photon energy.   The two components have broadly similar 
rms--energy shapes, although their relative strengths evolve during the
observed sequence. Although the shoulder remains broad, with
$Q\sim2$--$4$, its lag and rms--energy behavior resemble those of the
type-B QPO detected shortly after our observations. This similarity raises the
interesting possibility that the shoulder is related to an earlier, broader stage of
the variability later seen as the type-B QPO.}

   \maketitle
%

\section{Introduction}
\label{sec:intro}

Black-hole X-ray binaries (BHXRBs) are systems in which a stellar-mass black hole accretes matter from a companion star. 
Most BHXRBs are transients, spending long periods of time in quiescence and occasionally undergoing outbursts that last for weeks to months. 
During these episodes, the accretion luminosity increases by several orders of magnitude and the source moves through different spectral-timing states, commonly described by a characteristic “q”-shaped track in the hardness-intensity diagram \citep{belloni2005evolution,remillard2006x}. 
These states are usually classified as the low-hard state (LHS), hard-intermediate state (HIMS), soft-intermediate state (SIMS), and high-soft state (HSS), although some sources undergo failed-transition outbursts and remain in the LHS, sometimes reaching the HIMS, throughout the event \citep{homan2005evolution, capitanio2009failed, alabarta2021failed}. 

A key timing feature during BHXRB outbursts is the presence of QPOs, which appear as narrow peaks in the PDS. 
Low-frequency QPOs (LFQPOs) in black-hole systems are commonly classified into type-A, type-B, and type-C QPOs according to their centroid frequency, quality factor, fractional rms amplitude, noise properties, and phase lag behavior \citep{wijnands1999complex, casella2005abc,motta2016quasi}. 
Type-C QPOs are especially important because they are strong, coherent, and closely linked to the LHS and HIMS (see \citealp{ingram2019review} for a review). 
The centroid frequency, fractional rms amplitude, and energy-dependent phase lags of these QPOs are widely used to probe the geometry and radiative response of the inner accretion flow \citep{qu2010energy, ingram2009low, ingram2011physical, ingram2016quasi, van2016inclination,karpouzas2021variable}.

GX~339$-$4 is one of the best-studied black-hole X-ray binaries for this purpose \citep[e.g.][]{homan2005evolution, belloni2005evolution, motta2009evolution, motta2011low}. 
During its outbursts, this source shows strong type-C QPOs whose frequency evolves with the source state \citep{motta2009evolution, motta2011low, altamirano2015evolution}. 
Previous timing studies of GX~339$-$4 have shown that type-C QPO phase lags depend on QPO frequency and photon energy \citep{nowak1999rossi, altamirano2015evolution, zhang2017evolution}. 
In particular, \citet{zhang2017evolution} analyzes the \textit{RXTE} observations \citep{bradt1993x} in the rising phase of the 2006/2007 outburst and found that the lag-energy spectra of the type-C QPO evolve systematically as the QPO frequency increases. 
At low QPO frequencies the lags increase approximately monotonically with energy, while at higher QPO frequencies the lag-energy spectra become more complex and show structure around the iron-line energy. 
These results suggest that the QPO lag contains information about both the hard Comptonized emission and the reflected emission from the accretion disc \citep{kotov2001x, miller2007relativistic, zhang2017evolution}.

More recent studies have extended this picture with new timing and spectral-timing analyses of GX~339$-$4. 
Using \textit{RXTE}, \citet{zhang2024systematic} carried out a systematic study of the high-frequency bump associated with type-C QPO and showed that this feature is closely related to the evolution of QPO in LHS and HIMS. 
During the 2021 outburst, \citet{jin2023quasi} studied QPO in GX~339$-$4 with \textit{Insight-HXMT}, providing a high-energy view of the evolution of QPOs. \citet{zhang2024evolution} further investigated the evolution of type-C QPOs in GX~339$-$4 and EXO~1846$-$031, including the possible energy dependence of the QPO frequency and the fractional rms energy spectrum. 
AstroSat observations of the same outburst also showed complex energy-dependent QPO rms and time-lag behavior during the intermediate state \citep{mondal2023evolution}. 
In addition, \citet{jin2025broad} studied the broad-band noise components in GX~339$-$4 during the 2021 outburst, showing that the non-QPO Lorentzian components also evolve systematically with source state. 
These works reinforce the need to treat different variability components separately when interpreting the frequency- and energy-dependent timing behavior of GX~339$-$4.

The phase lag of a QPO is usually measured by averaging the real and imaginary parts of the CS over a finite interval around the QPO centroid, often one FWHM wide \citep[e.g.][]{reig2000phase, belloni2020time, zhu2023timing, zhu2024energy}. 
This gives a meaningful QPO lag when the QPO is the dominant contribution to the CS in that interval. 
In many black-hole binaries, however, the QPO sits on top of several broad and narrow Lorentzian components \citep{nowak2000there, belloni2002unified}. 
If one of these neighboring components has a different cross-spectral phase, the measured lag is the phase of the summed cross-vector, not the intrinsic phase lag of the QPO alone \citep{vaughan1997x, nowak1999rossi}. 
This becomes important when the QPO profile is asymmetric, when a shoulder is present close to the QPO peak, or when a weak component is more apparent in the CS than in the PDS  \citep[][hereafter M24]{mendez2024unveiling}.

To separate such blended components, M24 fitted the PDS and the real and imaginary parts of the CS simultaneously. 
In their model, each Lorentzian has the same centroid frequency and width in the PDS and in the CS, while its cross-spectral normalization and phase lag are fitted independently. 
This decomposition separates nearby components in the complex Fourier plane and can reveal components that are weak in the PDS but significant in the real or imaginary part of the CS \citep{fogantini2025hidden, bellavita2025nature}.

The \textit{RXTE} observation 92035-01-03-06 of GX~339$-$4 provides a direct example of this problem. 
Using the joint PDS$-$CS method, M24 showed that the type-C QPO region in this observation is better described by two nearby components: a narrow QPO fundamental and a broader high-frequency shoulder. 
The two components are close in Fourier frequency, but their phase lags are very different. 
The QPO lag is close to zero, whereas the shoulder has a much larger hard lag. 
This result showed that, within the joint PDS--CS decomposition, the
asymmetric QPO region can be represented by two Lorentzian contributions
with markedly different fitted phases. 

We therefore revisited the \textit{RXTE} observations of GX~339$-$4 during the rising phase of the 2006/2007 outburst, focusing on the observations analyzed by \citet{zhang2017evolution}. 
Several of their lag-frequency spectra show local structures around the type-C QPO region that resemble the shoulder feature reported by M24. 
This raised the possibility that the shoulder is not confined to a single observation, but appears and evolves over a broader part of the outburst. 
Since such a component can bias the lag measured by averaging over the QPO region, it is necessary to determine when the shoulder is present, how its frequency evolves relative to the QPO fundamental, and whether it has a distinct energy-dependent lag.

The aim of this work is to determine how the high-frequency shoulder is
represented in the PDS and CS, how its fitted properties evolve relative
to those of the type-C QPO, and how it affects phase lags measured over
the conventional QPO frequency interval.
In this work, we apply the joint PDS$-$CS fitting method of M24 to the full sequence of \textit{RXTE} observations in which a type-C QPO is detected during the 2006/2007 outburst. 
We use this decomposition to follow the emergence of the high-frequency QPO shoulder, to measure its phase lag separately from that of the QPO fundamental, and to compare their lag-energy spectra. 
The paper is organized as follows. 
Section~\ref{obs} describes the observations and data reduction. 
Section~\ref{Method} summarizes the joint PDS$-$CS fitting method. 
Section~\ref{Results} presents the QPO and shoulder evolution and their energy-dependent lags. 
Section~\ref{DISCUSSION} discusses the implications for QPO lag measurements and the nature of the shoulder. 
Section~\ref{conclusion} summarizes our conclusions. 
\section{Observations and data reduction}
\label{obs}

In this work, we revisit the \textit{RXTE} observations of GX~339$-$4 analyzed by \citet{zhang2017evolution}. Our sample consists of the same 23 public observations obtained during the rising phase of the 2006/2007 outburst, all of which were reported to show type-C QPOs. To place these observations in the broader context of the outburst, we show in Figure~\ref{afig:lc} the long-term evolution of the source during the 2006/2007 outburst. 
The data used to construct this figure are taken from \cite{motta2009evolution}, where a more detailed discussion of the outburst evolution is provided.

We used GHATS\footnote{\url{https://github.com/ghats-timing/ghats}} to compute the Fast Fourier transforms (FFTs) of the light curves in each energy band of interest, using 128~s segments and a time resolution of $1/1024$~s, corresponding to a Nyquist frequency of 512~Hz. PDS were computed in the full energy band and averaged over all segments, then geometrically rebinned in frequency by a factor of $\approx 1.023\,(=10^{1/100})$ to improve the signal-to-noise ratio while maintaining adequate frequency resolution. The PDS were normalized to fractional rms units  \citep{belloni1990atlas}, and the Poisson noise level was subtracted.
To examine whether the asymmetric QPO profile could be produced by intra-observation frequency drift, we additionally calculated a full-band dynamical PDS. We divided each continuous Good Time Interval (GTI) into 128-s segments, discarded segments
crossing data gaps, and rebinned the resulting PDS in frequency. For
display, the valid segments were concatenated and the GTI gaps were
omitted,  and the time axis of the dynamical PDS
therefore represents accumulated good exposure time.

For the main two-band CS analysis, we computed the PDS in the soft (2$-$5.7 keV) and hard bands (5.7$-$115 keV), together with the CS between these two bands. 
For comparison and for measuring the component frequencies and fractional rms amplitudes, we also computed the full-band PDS. 
The real and imaginary parts of the CS were averaged over all segments and rebinned in frequency in the same manner as the PDS for the subsequent joint timing analysis. The correspondence between energy bands and detector channels can be found in the \textit{RXTE}/Proportional Counter Array (PCA) Energy$-$Channel Conversion Table\footnote{\url{https://heasarc.gsfc.nasa.gov/docs/xte/e-c_table.html}}.

To investigate the energy dependence of the timing properties, we subdivided the PCA energy range into six bands: 2.0$-$5.7~keV, 5.7$-$7.7~keV, 7.7$-$10.6~keV, 10.6$-$15.0~keV, 15.0$-$20.6~keV, and 20.6$-$44.0~keV. These bands follow the broad PCA binning used by \citet{zhang2017evolution} for their rms-energy analysis. For the lag-energy analysis in this work, we used the 2.0$-$5.7~keV band as the reference band. This choice keeps the energy-resolved fits on the same soft-band reference system as the main two-band CS analysis and gives a high count rate in the reference band, which helps to stabilize the CS measurements of the relatively weak shoulder component. Energies above 44~keV were not used because of the low signal-to-noise ratio.

Using the same timing setup described above, we computed the PDS in each individual energy band and the CS between each higher-energy band and the 2.0$-$5.7~keV reference band. Guided by the centroid frequencies of the QPO components and by the frequency range over which the PDS and CS retain useful signal-to-noise, we restricted all fits to 0.01$-$20~Hz. This range includes the QPO fundamental, the shoulder, and the QPO harmonic components in all observations, while excluding the high-frequency regime where the PDS and CS are dominated by statistical noise.

\section{Method Overview}
\label{Method}

A phase lag is commonly estimated by averaging the real and imaginary parts of the CS over a narrow Fourier-frequency interval around the component of interest, usually one FWHM wide, and computing $\Delta\phi = \tan^{-1}(\langle \mathrm{Im} \rangle / \langle \mathrm{Re} \rangle)$ \citep{reig2000phase,belloni2020time}.
This estimate represents the lag of that component only if it dominates the CS within the chosen interval. 
When two nearby components contribute with different phases, the measured lag is instead the phase of their summed cross-vector, with weights set by their CS amplitudes (M24).

We therefore use the joint PDS$-$CS decomposition introduced by M24. 
The method assumes that the observed variability can be described as a sum of components that are coherent between energy bands but mutually incoherent, and that each component has its own transfer function $H_i(\nu)$ and phase lag $\Delta\phi_i(\nu)$. In this work we use the constant phase lag prescription, $\Delta\phi_i(\nu) = 2\pi k_i$. This is one of the simple forms considered by M24. It is adequate for the present fits because we model relatively narrow QPO-related features over a limited frequency range, where no phase wrapping is expected. With this prescription, the real and imaginary parts of the CS for component $i$ are proportional to $L_i\cos(2\pi k_i)$ and $L_i\sin(2\pi k_i)$, so $2\pi k_i$ is the fitted phase lag of that Lorentzian.

The joint decomposition is useful here because the shoulder can overlap the QPO fundamental in Fourier frequency while having a different phase. 
A fit to the PDS alone measures only the power-spectral shape, whereas the CS also constrains the direction of each component in the complex plane. 
A component that is weak in the PDS can therefore be required by the imaginary part of the CS or by the local phase lag spectrum.

We performed the fits in XSPEC v12.15.1 \citep{arnaud1996xspec}. 
Before fitting, we rotated the cross spectra counterclockwise by $45^\circ$ \citep{ fogantini2025hidden}. 
This brings the two fitted CS projections to comparable amplitudes when the phase lag is small. 
The rotation is linear and does not change the recovered component parameters.

For each observation, we first fitted the full-band PDS with a sum of Lorentzians. 
We began with broad Lorentzians for the underlying broadband noise and then added narrower components associated with the type-C QPO, the shoulder when present, and the harmonic structure. 
Components were added until the fit was acceptable and no coherent residuals remained around the QPO region. 
Except for the transitional shoulder-like component discussed below, the centroid frequencies and fractional rms amplitudes in Table~\ref{tab1} are taken from these full-band PDS fits.

We then used the full-band PDS decomposition as the component template for the joint fits. 
In the joint fits, the soft- and hard-band PDS and the real and imaginary parts of their CS were fitted simultaneously. 
For each Lorentzian, $\nu_{0,i}$ and $\Delta_i$ were tied across the two PDS and the two CS projections, while the PDS normalizations, CS normalizations, and phase lags were free. 
The phase lags reported in Table~\ref{tab1} are obtained from these joint PDS$-$CS fits. 

Additional Lorentzians were included in the joint fits only when they removed structured residuals in the CS and improved the overall fit without changing the neighboring components in an unstable way. 
When such a component was weak or not clearly resolved in the full-band PDS but required by the real or imaginary part of the CS, we classified it as a CS-selected component \citep{jin2025timing,rout2025hidden,fogantini2025hidden,jin2026black}.
\section{Results}
\label{Results}

\subsection{QPO fitting}
\begin{figure*}[htbp]
\centering
    \includegraphics[width=0.95\textwidth]{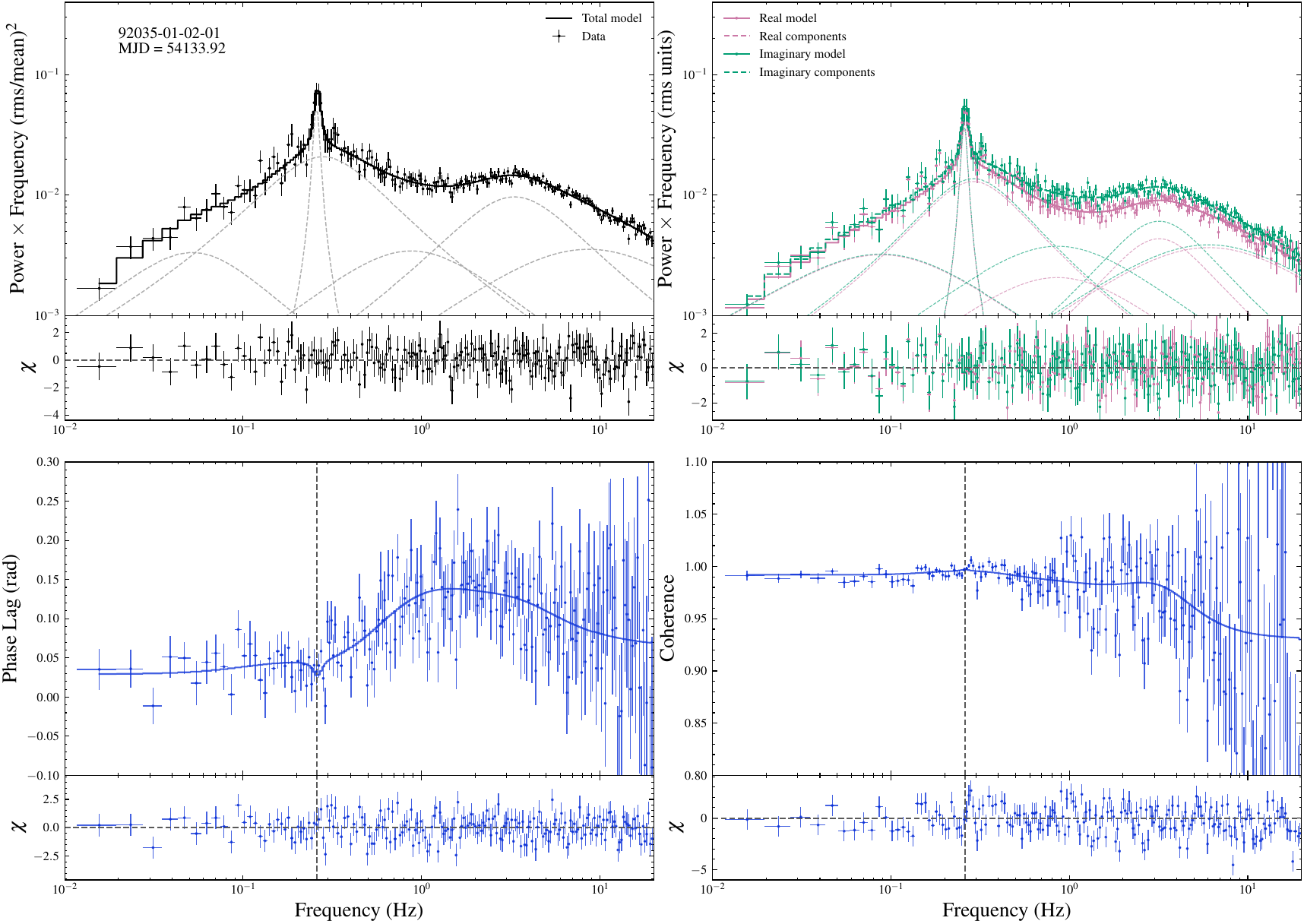}
\caption{
Joint fit to Obs 6 of GX 339$-$4 at MJD 54133.92. 
The upper-left panel shows the full-band PDS and its Lorentzian decomposition. 
The upper-right panel shows the real and imaginary parts of the CS between the two selected energy bands, with a rotation of $45^\circ$. 
The lower panels show the phase lag spectrum and intrinsic coherence derived (not fitted to these data) from the best-fitting PDS$-$CS model. 
Solid curves show the total model, dashed curves show individual Lorentzian components, and the lower sub-panels show the residuals in units of $\chi$. 
The vertical dashed line marks the centroid frequency of the type-C QPO. The full-band PDS is shown for decomposition reference; the joint fit was performed using the soft- and hard-band PDS and their CS.
}
    \label{fig:qpofit1}
\end{figure*}

Fig.~\ref{fig:qpofit1} shows a representative early observation, Obs 6 at MJD 54133.92. 
The full-band PDS is shown to illustrate the Lorentzian decomposition used to identify the variability components and to measure the centroid frequencies and fractional rms amplitudes. 
The joint fit itself was performed with the soft- and hard-band PDS, together with the real and imaginary parts of their CS. 
The corresponding soft- and hard-band PDS decompositions are shown in Appendix~\ref{app:fits} as Fig.~\ref{afig:fit1}. 

This observation contains a prominent type-C QPO on top of a broad low-frequency noise continuum. 
The same set of characteristic frequencies and widths describes the PDS and the two CS projections. 
The phase lag spectrum shows a small positive lag at the QPO frequency and a gradual increase toward higher frequencies. 
The intrinsic coherence remains close to unity around the QPO and becomes less constrained at higher frequencies, where the signal-to-noise ratio decreases. 
No additional high-frequency shoulder component is required at this stage. Notice that the phase lag spectrum and the coherence function are not fitted, but predicted from the fits to the PDS and CS. 

\begin{figure}
    \centering
    \includegraphics[width=0.85\columnwidth]{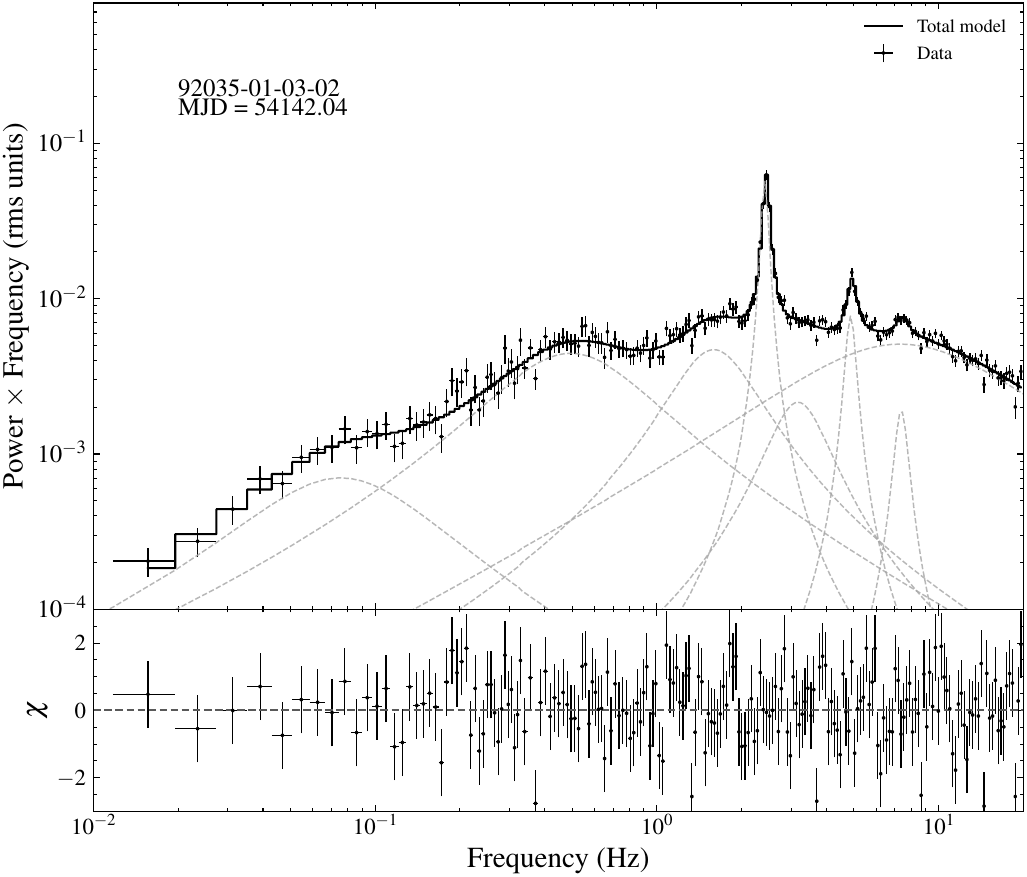}
\caption{
Full-band PDS of Obs 15 of GX 339$-$4 at MJD 54142.04.
The black solid curve shows the best-fitting total model, and the grey-dashed curves show the individual Lorentzian components.
}
    \label{fig:qpofit2}
\end{figure}

The first shoulder-like detection occurs in Obs 15 at MJD 54142.04. 
The full-band PDS of this observation is shown in Fig.~\ref{fig:qpofit2}. 
It contains a narrow type-C QPO at $\nu_{\rm QPO}\simeq2.45$ Hz, together with broad low-frequency noise and neighboring high-frequency structures. 
An eight-Lorentzian model gives an acceptable description of the full-band PDS, and the PDS residuals show no clear unresolved structure around the QPO frequency. 
From the PDS alone, an additional Lorentzian is therefore not required.

\begin{figure*}[htbp]
\centering
    \includegraphics[width=0.9\textwidth]{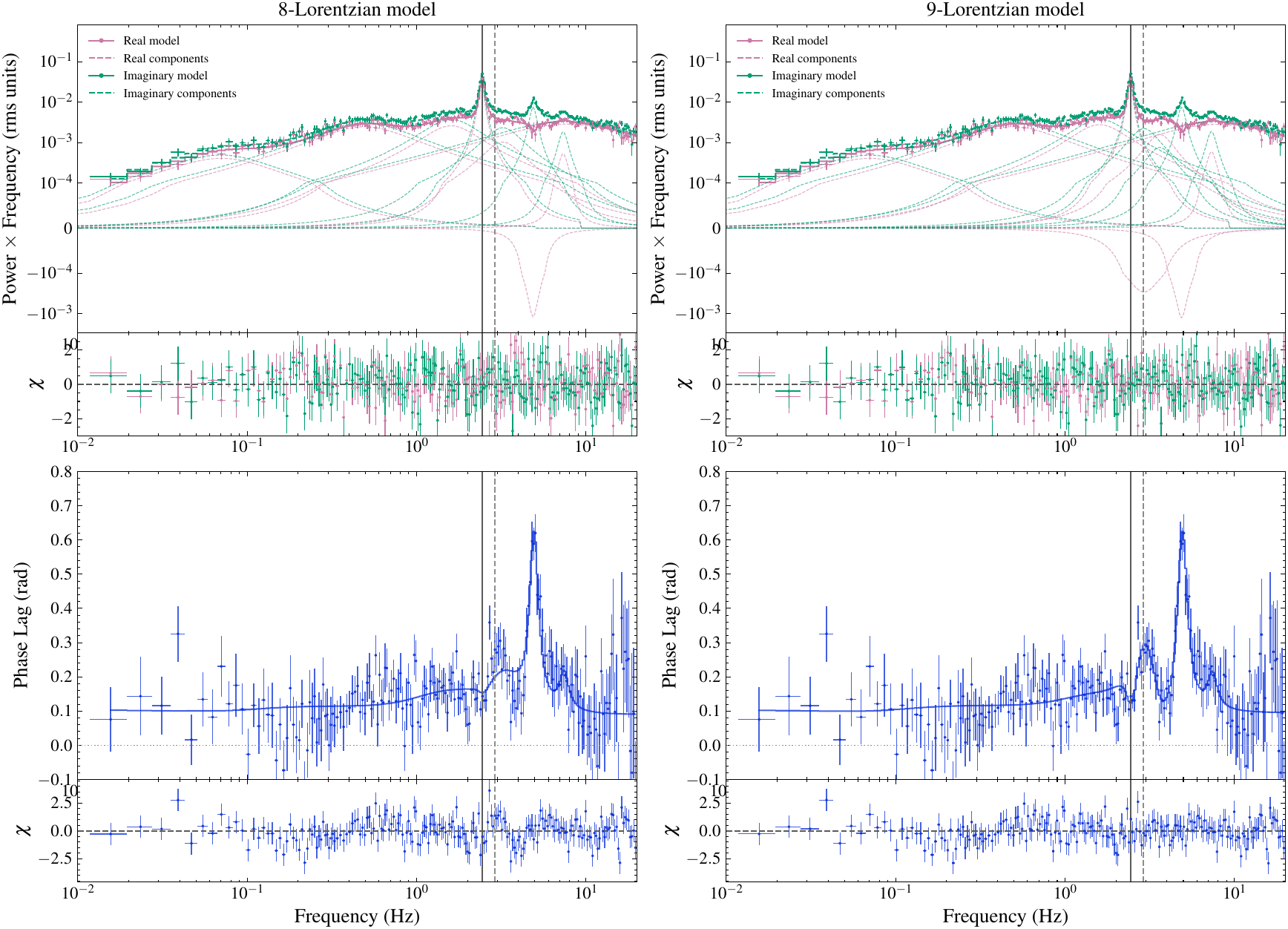}

\caption{
Comparison of the 8- and 9-Lorentzian joint fits for Obs 15 of GX 339$-$4 at MJD 54142.04.
The upper panels show the real and imaginary parts of the CS, and the lower panels show the corresponding phase lag spectra. 
The left and right columns show the 8- and 9-Lorentzian models, respectively. 
Solid curves show the total model, dashed curves show individual Lorentzian components, and the lower sub-panels show the residuals. 
The vertical dashed lines mark the centroid frequencies of the QPO-related components. 
Adding the ninth Lorentzian improves the fit from $\chi^2=818.8$ for 752 dof to $\chi^2=780.0$ for 746 dof. 
The improvement is mainly in the imaginary part of the CS and in the local phase lag structure around the QPO region.
}
    \label{fig:qpofit3}
\end{figure*}

Within the adopted multi-Lorentzian framework, the CS of the same
observation is better described when an additional Lorentzian is
included. Fig.~\ref{fig:qpofit3} compares the 8- and 9-Lorentzian joint fits to the real and imaginary parts of the CS and to the derived phase lag spectrum. 
The corresponding soft- and hard-band PDS and coherence spectra are shown in Appendix~\ref{app:fits} as Figs.~\ref{afig:fit2} and~\ref{afig:fit3}. 
The 8-Lorentzian model leaves a structured residual in the imaginary part of the CS around the QPO region, close to the narrow local enhancement in the phase lag spectrum. 
Adding a ninth Lorentzian reduces this residual and gives $\chi^2=780.0$ for 746 dof, compared with $\chi^2=818.8$ for 752 dof for the 8-Lorentzian model. 
Adding a tenth Lorentzian does not significantly improve the fit and does not appreciably change the QPO$-$shoulder parameters.

We therefore adopt the 9-Lorentzian model for this observation. 
The intrinsic coherence function, $\gamma^2$, of Obs.~15 is shown in
Figs.~\ref{afig:fit2} and \ref{afig:fit3}. The coherence is high
($\gamma^2 \approx 1$) at the QPO frequency. With increasing Fourier
frequency, it first decreases to $\gamma^2 \approx 0.75$ and then increases
again to $\gamma^2 \approx 0.9$, before decreasing to $\gamma^2 \approx 0.6$ at the frequency of the second
harmonic of the QPO. At even higher
frequencies, the coherence becomes increasingly uncertain because of the
declining signal-to-noise ratio. Although the coherence function is not
fitted directly, the coherence predicted by the 8-Lorentzian model does not
adequately reproduce the local evolution around the QPO frequency. In
contrast, the 9-Lorentzian model, which includes the additional
shoulder-like component, provides a better description of the observed
coherence structure at the shoulder frequency, marked by the vertical
dashed line.
The additional component is classified as CS-selected because the evidence for it comes mainly from the imaginary CS and the local phase lag feature, rather than from a distinct residual in the full-band PDS. 
We treat this observation as a transitional case: the QPO-region CS is
better described by an additional fitted phase contribution, while the
corresponding shoulder is not yet separately resolved in the full-band
PDS.

\begin{figure*}[htbp]
\centering
    \includegraphics[width=0.95\textwidth]{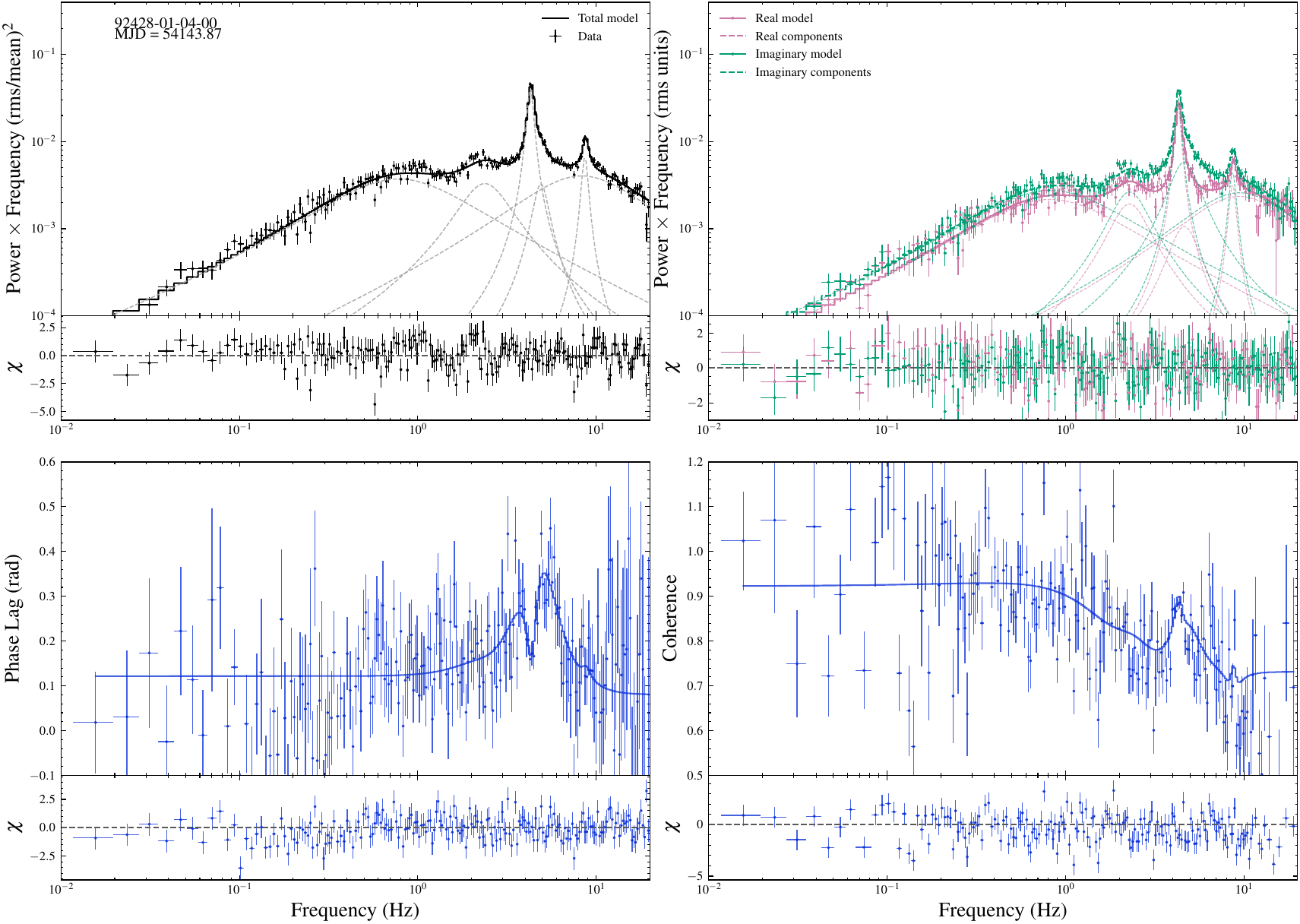}
\caption{Joint fit to Obs 17 of GX 339$-$4 at MJD 54143.87. 
The upper-left panel shows the full-band PDS and its Lorentzian decomposition. 
The upper-right panel shows the real and imaginary parts of the CS. 
The lower panels show the phase lag spectrum and intrinsic coherence derived from the best-fitting PDS$-$CS model. 
Solid curves show the total model, dashed curves show individual Lorentzian components, and the lower sub-panels show the residuals. 
At this epoch the shoulder is significant in the PDS and also contributes to the imaginary CS and the phase lag spectrum.
}
\label{fig:qpofit4}
\end{figure*}

The next observation with a shoulder detection, Obs 17 at MJD 54143.87, is shown in Fig.~\ref{fig:qpofit4}. 
The corresponding soft- and hard-band PDS are shown in Appendix~\ref{app:fits} as Fig.~\ref{afig:fit4}. 
In this case, the component on the high-frequency side of the QPO is visible in the full-band PDS and is included in the Lorentzian decomposition. 
The same component also produces a local enhancement in the imaginary part of the CS and in the phase lag spectrum. 
This observation therefore marks the stage at which the shoulder is no longer only CS-selected, but becomes a resolved timing component in the PDS as well.
The best-fitting QPO and shoulder parameters are summarized in Table~\ref{tab1}. We also show the temporal evolution of the quality factor of the shoulder in Fig.~\ref{q}.

\begin{figure}[htbp]
\centering
\includegraphics[width=0.95\columnwidth]{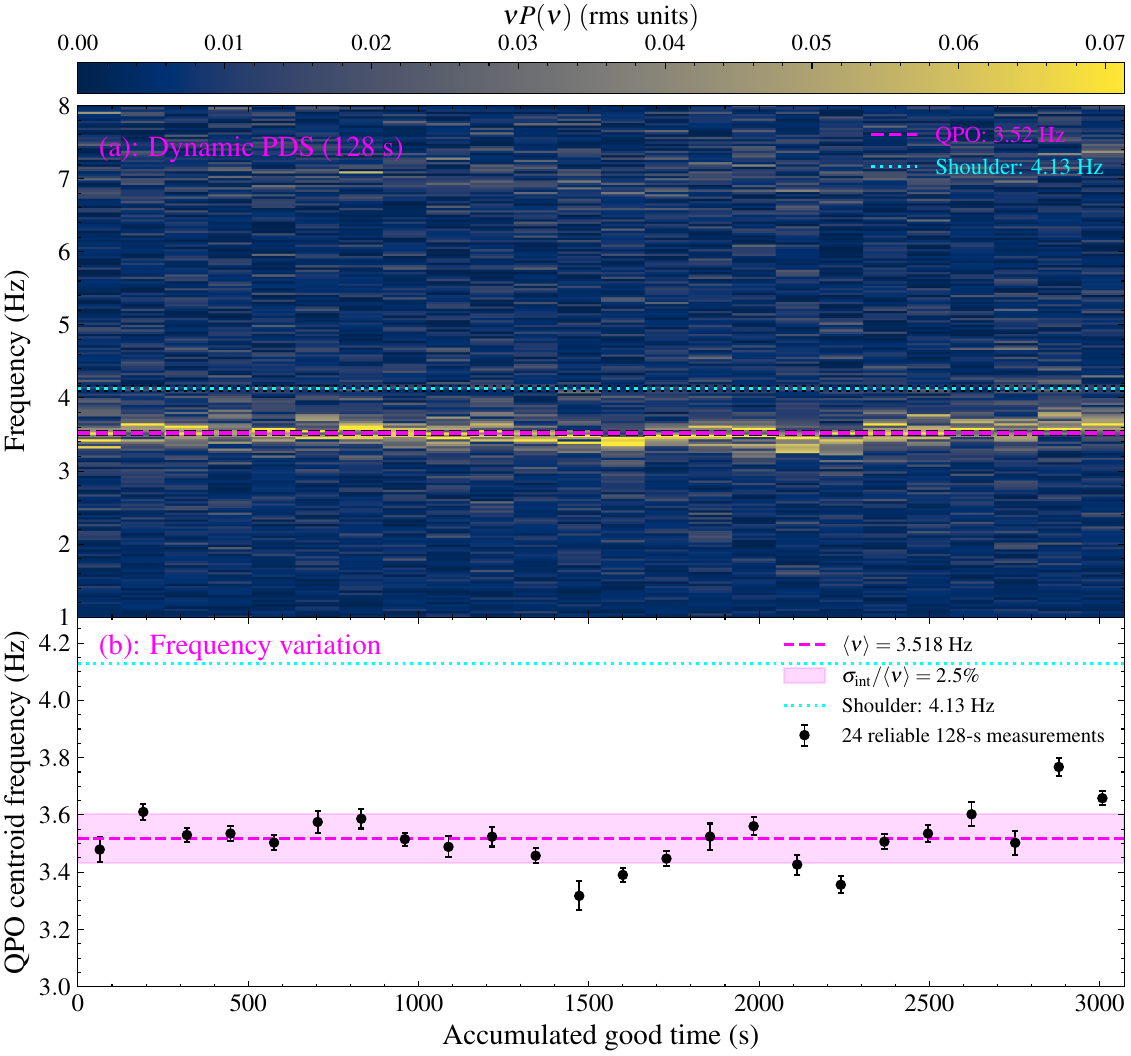}
\caption{
Time-resolved variability of the QPO in Obs.~16. The upper
panel shows the 128-s dynamical PDS. The magenta dashed and cyan dotted lines
indicate the time-averaged QPO centroid frequency
($3.52~\mathrm{Hz}$) and the shoulder frequency
($4.13~\mathrm{Hz}$), respectively. The lower panel shows the QPO
centroid frequencies obtained from local Lorentzian fits to all 24
intervals. The magenta dashed line marks the mean centroid frequency,
$\langle\nu\rangle=3.518~\mathrm{Hz}$, while the shaded region
represents the intrinsic scatter,
$\sigma_{\rm int}/\langle\nu\rangle\simeq2.5\%$. The GTI gaps have been
omitted, and the horizontal axis therefore represents accumulated good
exposure time. The measured frequency variability is substantially
smaller than the separation between the QPO and the shoulder.}
\label{fig:qpo_frequency_variation}
\label{fig:dynpds}
\end{figure}

\begin{table*}[htbp]
    \centering
\caption{
Best-fitting parameters of the type-C QPO fundamental and the QPO shoulder of GX 339$-$4 for the selected \textit{RXTE} observations.
}
    \label{tab1}
    \renewcommand{\arraystretch}{1.1}
    \setlength{\tabcolsep}{5pt}

\begin{threeparttable}
    \begin{tabular}{ccccccccc}
        \toprule \toprule
       No.& ObsID & MJD & $\nu$ QPO & Phase lag & $rms$ & $\nu$ \textit{shoulder} & Phase lag \textit{shoulder} & $rms$ \\
        & & & (Hz) & (rad) & (\%) & (Hz) & (rad) & (\%) \\
        \midrule
1&92035-01-01-01 & 54128.94 & $0.14\pm0.01$ & $0.02\pm0.10$ & $12.3\pm1.0$ & -- & -- & -- \\
2&92035-01-01-03 & 54130.13 & $0.16\pm0.01$ & $0.01\pm0.13$ & $11.4\pm1.8$ & -- & -- & -- \\
3&92035-01-01-02 & 54131.11 & $0.18\pm0.01$ & $0.02\pm0.09$ & $9.8\pm1.3$ & -- & -- & -- \\
4&92035-01-01-04 & 54132.09 & $0.20\pm0.01$ & $0.02\pm0.11$ & $9.2\pm1.2$ & -- & -- & -- \\
5&92035-01-02-00 & 54133.00 & $0.23\pm0.01$ & $0.01\pm0.09$ & $9.7\pm1.1$ & -- & -- & -- \\
6&92035-01-02-01 & 54133.92 & $0.26\pm0.01$ & $0.02\pm0.12$ & $8.5\pm1.1$ & -- & -- & -- \\
7&92035-01-02-02 & 54135.03 & $0.29\pm0.01$ & $0.02\pm0.06$ & $10.0\pm1.3$ & -- & -- & -- \\
8&92035-01-02-03 & 54136.02 & $0.36\pm0.01$ & $0.02\pm0.07$ & $11.0\pm0.7$ & -- & -- & -- \\
9&92035-01-02-04 & 54137.00 & $0.42\pm0.01$ & $0.03\pm0.06$ & $10.4\pm1.1$ & -- & -- & -- \\
10&92035-01-02-08 & 54137.85 & $0.54\pm0.01$ & $0.01\pm0.09$ & $10.0\pm1.4$ & -- & -- & -- \\
11&92035-01-02-07 & 54138.83 & $0.89\pm0.01$ & $0.10\pm0.05$ & $9.84\pm0.48$ & -- & -- & -- \\
12&92035-01-02-06 & 54139.94 & $0.99\pm0.01$ & $0.17\pm0.06$ & $4.33\pm0.38$ & -- & -- & -- \\
13&92035-01-03-00 & 54140.20 & $1.13\pm0.01$ & $0.14\pm0.05$ & $4.97\pm0.22$ & -- & -- & -- \\
14&92035-01-03-01 & 54141.06 & $1.68\pm0.01$ & $0.14\pm0.04$ & $6.25\pm0.19$ & -- & -- & -- \\
\hline
15&92035-01-03-02\tnote{a} & 54142.04 & $2.45\pm0.01$ & $0.11\pm0.03$ & $7.31\pm0.18$ & $2.90\pm0.09$ & $0.92\pm0.28$ & $3.7\pm0.9$ \\
\hline
16&92035-01-03-03 & 54143.02 & $3.52\pm0.01$ & $0.15\pm0.03$ & $7.34\pm0.26$ & $4.13\pm0.12$ & $0.71\pm0.10$ & $3.3\pm0.6$ \\
17&92428-01-04-00 & 54143.87 & $4.32\pm0.01$ & $0.12\pm0.03$ & $7.34\pm0.24$ & $4.93\pm0.17$ & $0.61\pm0.07$ & $3.2\pm0.9$ \\
18&92428-01-04-01 & 54143.95 & $4.22\pm0.02$ & $0.11\pm0.04$ & $7.07\pm0.26$ & $4.6\pm0.6$ & $0.79\pm0.11$ & $5.0\pm1.6$ \\
19&92428-01-04-02 & 54144.09 & $4.13\pm0.01$ & $0.14\pm0.05$ & $7.19\pm0.26$ & $4.27\pm0.12$ & $0.66\pm0.11$ & $5.0\pm1.0$ \\
20&92428-01-04-03 & 54144.87 & $4.96\pm0.01$ & $0.13\pm0.05$ & $6.03\pm0.64$ & $5.17\pm0.15$ & $0.53\pm0.05$ & $5.1\pm1.0$ \\
21&92035-01-03-05 & 54145.11 & $5.72\pm0.01$ & $-0.02\pm0.06$ & $4.9\pm1.1$ & $6.1\pm1.1$ & $0.56\pm0.06$ & $5.7\pm1.8$ \\
22&92428-01-04-04 & 54145.97 & $5.58\pm0.02$ & $0.05\pm0.05$ & $5.49\pm0.51$ & $6.24\pm0.42$ & $0.53\pm0.08$ & $3.6\pm1.5$ \\
23&92035-01-03-06 & 54146.03 & $5.48\pm0.01$ & $0.03\pm0.05$ & $4.52\pm0.48$ & $5.83\pm0.12$ & $0.55\pm0.05$ & $5.5\pm0.7$ \\
        \bottomrule
    \end{tabular}
\tablefoot{ For each observation, we list the centroid frequency, phase lag, and fractional rms amplitude of the QPO fundamental; when a shoulder component is required, the corresponding shoulder parameters are also reported.
The centroid frequencies and fractional rms amplitudes are obtained from the full-band PDS fits.
The phase lags are measured from the joint fits to the soft- and hard-band PDS and to the real and imaginary parts of their CS, using the same Lorentzian component assignment as in the full-band PDS decomposition.  \tablefoottext{a}{For this observation, the shoulder component is not independently required by the full-band PDS, but is required by the imaginary part of the CS and by the narrow local feature in the derived phase lag spectrum. 
We therefore regard it as a transitional case, in which the QPO-region CS phase structure has already become more complex while the shoulder is not yet clearly separated in the PDS. 
Because the shoulder component is not detected in the full-band PDS in this observation, its centroid frequency, phase lag, and fractional rms amplitude are taken from the joint PDS--CS fit.}}

    \end{threeparttable}
\end{table*}

Although the 128-s dynamical PDS shows short-timescale variations in
the QPO centroid, there is no systematic shift from
$\sim3.5~\mathrm{Hz}$ toward the shoulder at $4.13~\mathrm{Hz}$
(Fig.~\ref{fig:dynpds}a). To quantify these variations, we fitted each
of the 24 individual 128-s PDS over a restricted frequency range around
the QPO using a local continuum plus a Lorentzian component. The QPO
centroid was reliably constrained, with finite two-sided uncertainties,
in all 24 intervals. After accounting for the statistical uncertainty
of each measurement, we obtain a mean centroid frequency of
$\langle\nu\rangle=3.518~\mathrm{Hz}$ and an intrinsic rms scatter of
$\sigma_{\rm int}\simeq0.088~\mathrm{Hz}$, corresponding to
$\sigma_{\rm int}/\langle\nu\rangle\simeq2.5\%$
(Fig.~\ref{fig:dynpds}b). This scatter is substantially smaller than
the $\sim0.61~\mathrm{Hz}$ separation between the mean QPO centroid and
the shoulder.

\subsection{Evolution of the QPO and shoulder parameters}

\begin{figure}[htbp]
\centering
    \includegraphics[width=0.95\columnwidth]{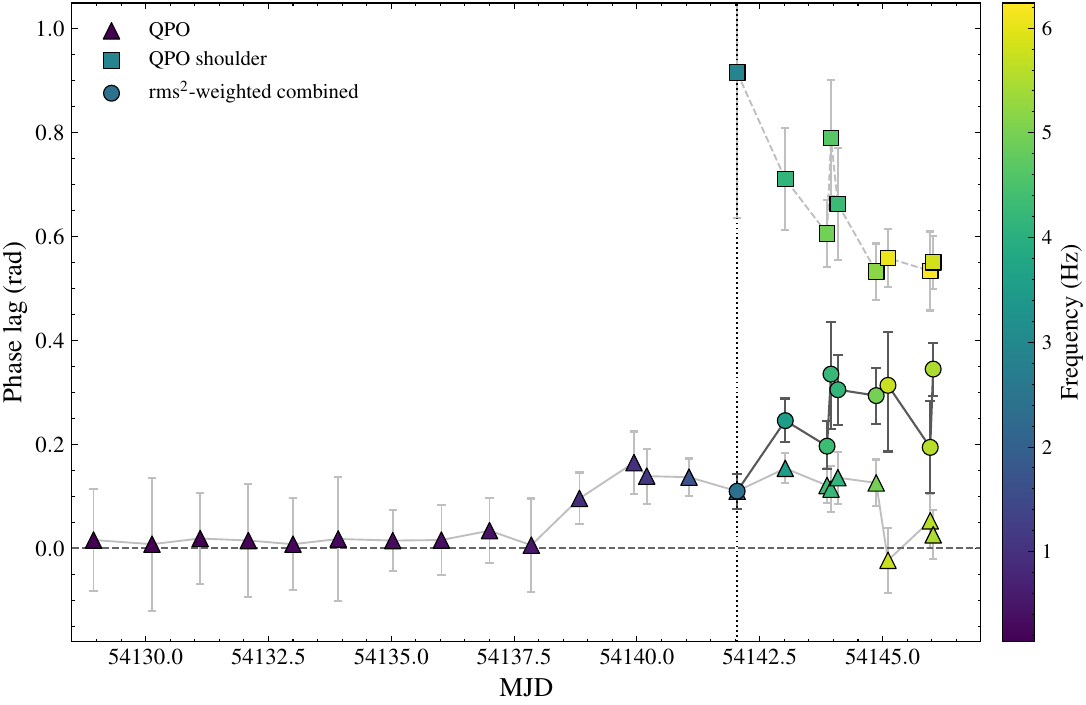}
\caption{
Evolution of the phase lags of the Type-C QPO and its high-frequency shoulder in GX~339$-$4 during the rising phase of the 2006/2007 outburst. Triangles and squares show the QPO and shoulder lags, respectively, while circles show their rms$^{2}$-weighted vector-combined lag. Colors indicate centroid frequency, using the QPO frequency for the combined points. The horizontal dashed line marks zero lag, and the vertical dotted line marks MJD~54142.04, when the shoulder is first identified in the CS. No combined lag is shown for this observation because the shoulder is not independently detected in the full-band PDS.}

    \label{fig:phase}
\end{figure}

Fig.~\ref{fig:phase} shows the phase lag evolution of the QPO fundamental and the shoulder. 
The QPO fundamental is detected in all selected observations. 
Its centroid frequency increases from $\sim0.14$ Hz to $\sim5$$-$$6$ Hz during the sequence, while its fractional rms amplitude decreases from $\sim10$$-$$12\%$ to $\sim4$$-$$7\%$. 
Its phase lag remains small: the early measurements are mostly consistent with zero within the 1$\sigma$ uncertainties, the lag reaches $\sim0.1$$-$$0.17$ rad around MJD 54139$-$54144, and then drops below $\sim0.1$ rad in the last observations.

The shoulder is detected from MJD 54142.04 onward. 
In the transitional observation, its centroid frequency is $2.90\pm0.09$ Hz, compared with $2.45\pm0.01$ Hz for the QPO fundamental. 
In later observations, the shoulder remains on the high-frequency side of the QPO and reaches $\sim5$$-$$6$ Hz. 
Its fractional rms amplitude is typically $\sim3$$-$$6\%$. 
Its phase lag is much larger than that of the QPO: the transitional observation gives $0.916\pm0.280$ rad, and the later detections give $\sim0.5$$-$$0.8$ rad.

To illustrate the combined contribution of the QPO and shoulder across the QPO frequency range, we calculated an rms$^{2}$-weighted vector-combined phase lag using their fitted phase lags and fractional rms amplitudes. For the transitional Obs.~15 at MJD~54142.04, we do not report the combined phase lag because the shoulder is not independently detected in the full-band PDS and its rms amplitude is inferred from the joint PDS--CS decomposition. In the subsequent observations, both the QPO and shoulder are resolved in the PDS, and the combined phase lag lies between their individual values, with typical values of $\sim0.2$--$0.35$~rad.

\begin{figure}[htbp]
\centering
    \includegraphics[width=0.9\columnwidth]{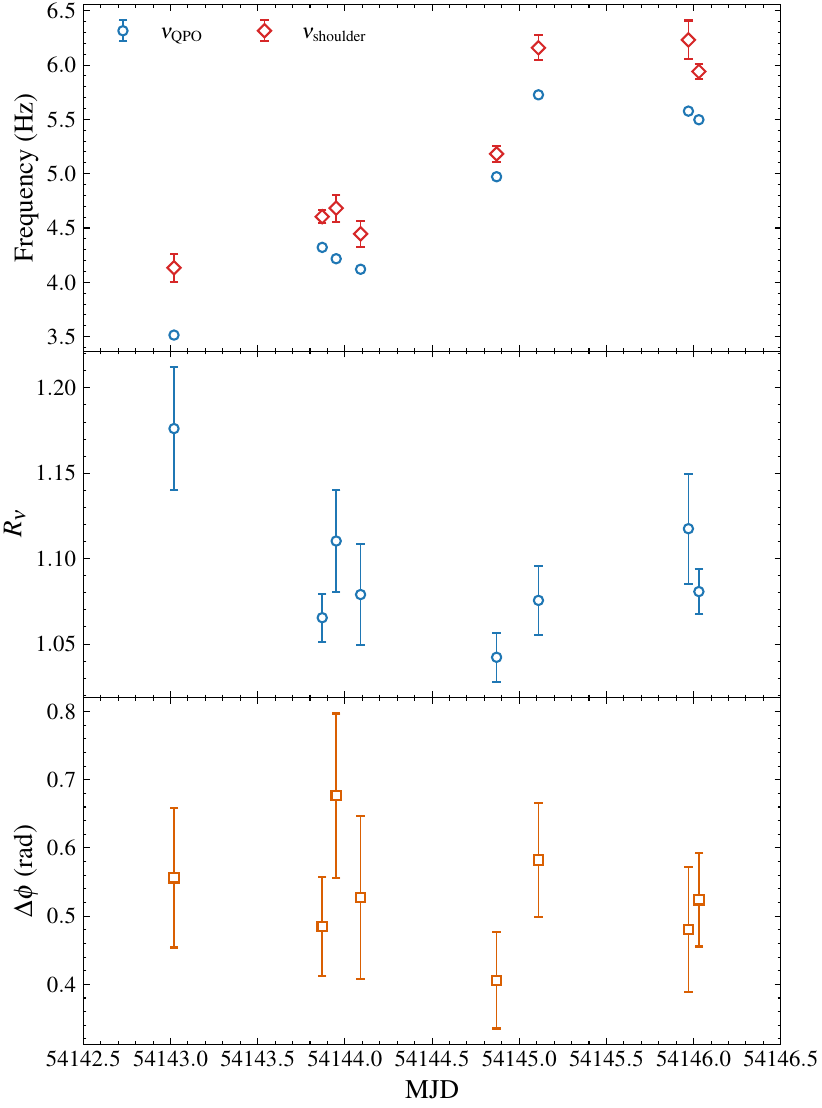}
\caption{
Evolution of the QPO fundamental and shoulder of GX~339$-$4 during the rising phase of the 2006/2007 outburst. 
From top to bottom, the panels show the centroid frequencies, frequency ratio $R_{\nu}$, phase lag difference $\Delta\phi$, and the lag separation in units of the propagated 1$\sigma$ uncertainty.   
}
    \label{fig:eve}
\end{figure}

Fig.~\ref{fig:eve} compares the centroid frequencies and phase lag differences of the two components for the observations with a shoulder detection. We define $R_{\nu}=\nu_{\rm sh}/\nu_{\rm QPO}$, where $\nu_{\rm QPO}$ and $\nu_{\rm sh}$ are the centroid frequencies of the QPO fundamental and the shoulder. We also define $\Delta\phi=\phi_{\rm sh}-\phi_{\rm QPO}$, where $\phi_{\rm QPO}$ and $\phi_{\rm sh}$ are the phase lags measured from the joint PDS$-$CS fits. The lag separation is expressed in units of the propagated quoted 1$\sigma$ uncertainty.

The shoulder tracks the QPO frequency evolution and remains on its high-frequency side. 
The frequency ratio is $R_{\nu}\simeq1.04$$-$$1.18$, far below the harmonic value of 2. 
The phase lag difference is positive in every detection, with 
$\Delta\phi\simeq0.4$$-$$0.7$ rad. 
Using the quoted 1$\sigma$ uncertainties, the lag separation exceeds $4\sigma$ in all cases.
Thus the shoulder is close to the QPO in frequency but has a distinct cross-spectral phase.

\subsection{Energy dependence}

We measured the energy dependence of the QPO and shoulder lags using the six PCA energy bands defined in Section~\ref{obs}. 
The 2.0$-$5.7 keV band was used as the reference band, and the lags were measured for the five higher-energy bands. 
For each observation and each energy band, we repeated the same PDS$-$CS fitting procedure. 
Using the same reference band as in the main two-band CS analysis keeps the full-band and energy-resolved fits on the same lag convention and gives a higher reference-band count rate for the weak shoulder component.

This reference band differs from that of \citet{zhang2017evolution}, who used 2$-$4 keV. 
The energy-dependent lags reported here are therefore intended for the component-resolved comparison between the QPO fundamental and the shoulder, rather than for a direct numerical comparison with their lag-energy spectra.

For the observations preceding the emergence of the shoulder, we obtained QPO lag$-$energy spectra with shapes similar to those reported by \citet{zhang2017evolution}. Since no additional shoulder component is required in these observations, we do not show these spectra separately and focus below on the observations in which the QPO and shoulder can be resolved.

\begin{figure*}[htbp]
\centering
    \includegraphics[scale=0.45]{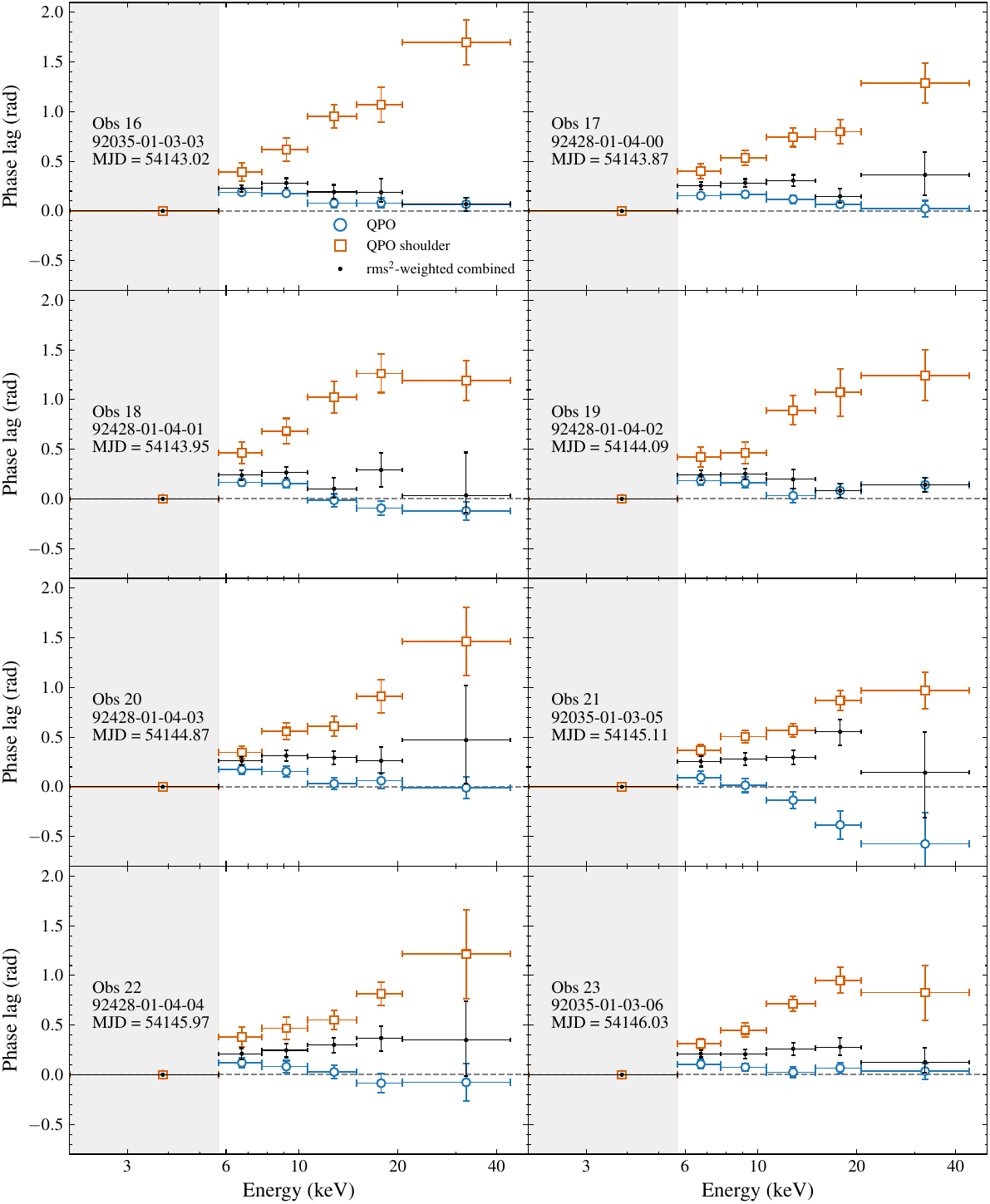}
    \includegraphics[scale=0.45]{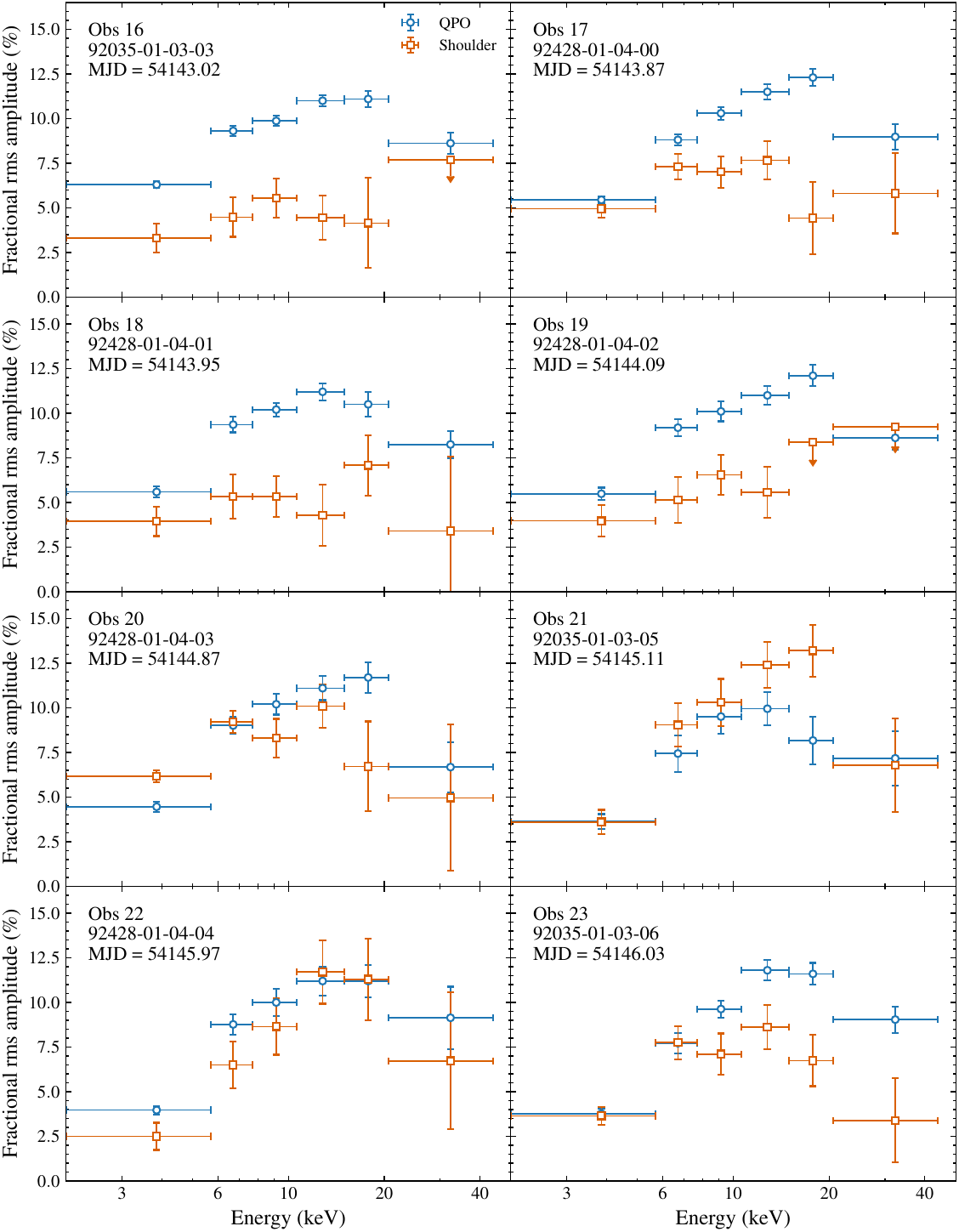}
\caption{Energy dependence of the phase lags (left) and fractional rms amplitudes
(right) of the QPO fundamental and the associated high-frequency
shoulder of GX~339$-$4 in observations where the shoulder is resolved as
a separate component. Each panel corresponds to one observation, with
the observation number, ObsID, and MJD labeled. Blue circles denote the
QPO fundamental, orange squares denote the shoulder, and black points
 show the rms$^{2}$-weighted vector-combined
phase lag calculated from the phase lags and fractional rms amplitudes
of the QPO and shoulder components. The vertical error bars indicate the
corresponding uncertainties, while the horizontal error bars show the
energy-bin widths. The shaded gray region indicates the 2.0--5.7~keV
reference band. In the lag panels, the lag of the reference band is set
to zero by convention, and the plotted values show the phase lags of the
higher-energy bands relative to this reference, as measured from the
joint PDS--CS decomposition. The dashed horizontal line marks zero lag.}
    \label{fig:lag_energy}
\end{figure*}

We exclude Obs 15 from the energy-resolved shoulder analysis. 
In that observation, the shoulder-like component is selected mainly by the CS and is not clearly resolved in the PDS. 
The resulting lag-energy spectra are shown in the left panel of Fig.~\ref{fig:lag_energy}.
In all observations, the shoulder lag is larger than the QPO lag over the PCA energy range. 
The QPO lag usually remains below $\sim0.2$ rad in the lower and intermediate energy bands, and it is consistent with zero or slightly negative at the highest energies in several observations. 
The shoulder instead shows an increasing hard lag, from $\sim0.3$$-$$0.5$ rad in the lowest non-reference band to values of order $\sim0.8$$-$$1.7$ rad at higher energies.

The largest shoulder lags occur in the 20.6$-$44.0 keV band, where the uncertainties are also largest because of the lower count rate. 
These high-energy points show that the shoulder remains more delayed than the QPO, but their exact numerical values are less robust than those in the lower-energy bands. 

In addition, we examined the energy dependence of the fractional rms amplitudes of the QPO fundamental and the shoulder. The resulting rms$-$energy spectra are shown in the right-hand panels of Fig.~\ref{fig:lag_energy}. The QPO rms generally increases with energy from the reference band to $\sim$10$-$20~keV, reaching typical amplitudes of $\sim$10$-$12\%, and then flattens or slightly decreases at the highest energies in several observations. The shoulder displays a more variable behavior. In the earliest observations where it is resolved as a separate component, its rms is lower than that of the QPO, with amplitudes ranging from only a few per cent to $\sim$7\%. During the subsequent evolution, the shoulder strengthens and becomes comparable to the QPO in the intermediate-energy bands, reaching $\sim$9$-$13\% around $\sim$8$-$20~keV in some observations. At the highest energies, the shoulder rms is less well constrained and shows larger scatter. 

\section{Discussion}
\label{DISCUSSION}
Using a joint multi-Lorentzian decomposition of the PDS and CS of
GX~339$-$4 during its 2006/2007 outburst, we find that the QPO region is
well described by a narrow type-C QPO and a neighboring high-frequency
shoulder. The shoulder is first identified mainly through the imaginary
part of the CS and the phase-lag spectrum, and later becomes visible as
a broader feature in the PDS. Within this decomposition, the shoulder carries a
larger hard lag and follows a different energy-dependent evolution from
the QPO fundamental.

The shoulder appears near the end of the HIMS, shortly before the
type-B QPO is detected. Its energy-dependent properties
therefore provide a possible link with the spectral-timing evolution
near the HIMS--SIMS transition.

\subsection{Cross-spectral characterization of the shoulder}

The shoulder is first identified in the adopted decomposition through
its cross-spectral behavior. 
Multi-Lorentzian decompositions are commonly used to describe the broad-band variability of accreting black holes, and the fitted components are usually interpreted as variability features with characteristic time-scales rather than as only mathematical terms \citep{nowak2000there,belloni2002unified}. 
In GX~339$-$4, earlier work already showed that different Lorentzian components can carry different phase lag and coherence properties \citep{nowak1999rossi}. 
The shoulder should therefore be assessed not only from its appearance in the PDS, but also from its contribution to the complex CS.

The transitional observation at MJD~54142.04 provides the first indication of a shoulder-like feature. The baseline eight-Lorentzian model gives an acceptable fit to the full-band PDS, with no clear residual around the QPO peak, but leaves a structured residual in the imaginary part of the CS and does not reproduce the narrow feature in the phase-lag spectrum. Adding a ninth Lorentzian improves the description of both features, so the evidence for this component at this stage comes mainly from the CS rather than from the full-band PDS.

By Obs.~17, a component on the high-frequency side of the QPO is also resolved in the PDS and remains visible in the imaginary CS and the phase-lag spectrum. The sequence therefore traces the development of the feature from a CS-selected component to a resolved power-spectral shoulder. 

The term QPO shoulder follows the usage of M24. 
In their analysis of Obs 23, the QPO profile of GX~339$-$4 was described by a narrow QPO component and a broader component on the high-frequency side of the peak. 
They noted that this broad component is reminiscent of the shoulder reported by \citet{belloni1997energy}, later referred to as the ``hump'' in the multi-Lorentzian classification of \citet{belloni2002unified}, and called it the QPO shoulder. 
The two components had close but distinct centroid frequencies, $\nu_{\rm QPO}=5.50\pm0.01$~Hz and $\nu_{\rm sh}=5.78\pm0.05$~Hz, and were also separated by their cross-spectral phases.

As expected from the use of the same joint PDS--CS method, our fit of this
observation gives values comparable to those of M24. We measure
$\phi_{\rm QPO}=0.026\pm0.047$ rad and
$\phi_{\rm sh}=0.551\pm0.051$ rad, compared with
$\phi_{\rm QPO}=-0.004\pm0.024$ rad and
$\phi_{\rm sh}=0.51\pm0.03$ rad in M24. In both analyses, the QPO lag is
close to zero, while the shoulder carries a much larger hard lag. The small
differences arise from the energy bands used, the different timing setups,
and the detailed Lorentzian decompositions.

The early behavior of the GX~339$-$4 shoulder is related to the recently reported hidden or imaginary QPOs. 
M24 showed that a component with a small real CS contribution but a large imaginary contribution can be difficult to identify in the PDS and can instead appear mainly through the CS. 
Similar cases have been reported in MAXI~J1820+070  \citep{bellavita2025nature} and Cyg~X$-$1 \citep{fogantini2025hidden}, where QPO-like signals are significant mainly in the imaginary part of the CS or in the associated lag and coherence spectra.  
The first shoulder detection in GX~339$-$4 resembles these cases because the additional Lorentzian is selected by the imaginary CS and by the local phase lag feature.

In GX~339$-$4, the shoulder component does not remain hidden. 
In the following observations it becomes visible as a high-frequency shoulder in the PDS, so it is not a purely imaginary QPO in the strict sense. 
We describe it as a CS-selected shoulder at its first appearance and as a resolved component of the type-C QPO complex once it becomes visible in the PDS. 
This description accounts for both its initial imaginary-CS detection and its later power-spectral visibility.

\subsection{Frequency proximity and phase lag separation}

In the full sequence, the shoulder lies on the high-frequency side of the QPO and follows the same overall frequency evolution. 
The frequency ratio remains close to unity, $R_{\nu}\simeq1.04$$-$$1.18$, placing it much closer to the QPO fundamental than to the harmonic. 
We therefore identify the shoulder as a neighboring component of the type-C QPO complex rather than as part of the harmonic sequence.

The near-unity frequency ratio suggests that the QPO fundamental and the shoulder are coupled to the same evolving time-scale in the accretion flow. 
Several Lorentzian components are known to move together as X-ray binaries change state \citep{psaltis1999correlations,belloni2002unified,pottschmidt2003long}. 
In GX~339$-$4, the type-C QPO frequency evolves rapidly during the 2006/2007 rising phase together with the spectral state and lag behavior \citep{motta2009evolution,motta2011low,zhang2017evolution,altamirano2015evolution}. 
A related result comes from \citet{zhang2024systematic}, who showed that the broad high-frequency bump in GX~339$-$4 also tracks the type-C QPO and source hardness, although at much higher characteristic frequencies than the shoulder studied here. 
Recent work on GX~339$-$4 and MAXI~J1348$-$630 also suggests that the QPO frequency is more closely linked to hardness than to luminosity \citep{zhang2024evolution,wang2026full}. 
The shoulder follows the QPO frequency evolution in this same sense, but remains slightly displaced to higher frequency.

The phase lag behavior separates the two components more strongly than their frequencies do. 
Type-C QPO lags are known to depend on source state, QPO frequency, inclination, and which Fourier component is measured \citep{van2016inclination,zhang2017evolution,zhang2020systematic,belloni2024fast}. 
In our decomposition, the QPO fundamental retains small lags throughout, while the shoulder carries a much larger hard lag. 
The difference $\Delta\phi=\phi_{\rm sh}-\phi_{\rm QPO}$ is positive in every detection, with typical values of $\sim0.4$$-$$0.7$ rad (see Fig.~\ref{fig:eve}).

The time-resolved fits confirm that the QPO is not strictly stationary,
with $\langle\nu\rangle=3.518~\mathrm{Hz}$ and an intrinsic rms scatter
of $\sigma_{\rm int}\simeq0.088~\mathrm{Hz}$, corresponding to only
$\sim2.5\%$ of the mean frequency (Fig.~\ref{fig:dynpds}b). This scatter
is approximately seven times smaller than the $\sim0.61~\mathrm{Hz}$
separation between the QPO and the shoulder. Thus, although temporal
frequency drift may contribute to the width or detailed shape of the
QPO profile, its measured amplitude is insufficient to account for the
shoulder and is unlikely to be its primary cause.

As a complementary test that is less sensitive to frequency drift over the full observation, we divided Obs.~16 into continuous 256-s intervals and calculated an averaged cross spectrum within each interval using eight 32-s Fourier segments. The short-interval cross spectra and results are presented in Fig.~\ref{app:short_cross}. The shoulder remains clearly visible in these short-interval cross spectra, particularly in the imaginary component, at a frequency consistent with that measured from the full observation. Its persistence when the contribution of longer-timescale temporal
frequency drift is reduced argues against the shoulder being produced
solely by averaging over the time evolution of the QPO centroid. This interpretation is also supported by the detection of the shoulder in the imaginary cross spectrum and by its substantially larger hard lag relative to the QPO fundamental. Together, these results argue against intra-observation temporal
frequency drift as the sole origin of the shoulder. They do not,
however, exclude an energy-dependent QPO centroid or more general
energy-dependent, nonstationary transfer-function effects.

Nevertheless, the dynamical PDS is subject to the usual trade-off between time and frequency resolution. Frequency variations  below the statistical sensitivity of the individual PDS, may remain unresolved and could still affect the detailed widths and shapes of the fitted components. A more detailed characterization of such variations would require complementary time-localized methods, such as wavelet analysis or the Hilbert--Huang transform \citep{chen2022wavelet,yu2023hilbert,shui_phase-resolved_2024,jin2024wavelet,zhu2026timing,zhu2026timingb}. A dedicated application of these methods to the shoulder is beyond the scope of the present work and will be presented elsewhere (Zhu et al., in preparation).

The coherence offers an additional view of the relation between the
variability in the two energy bands. A coherence close to unity is expected
when the variability is produced by a common driving signal acting upon two
different deterministic transfer functions. A decrease in coherence may
indicate that more than one contribution is present if components with
different phase lags or energy dependence overlap in frequency, since their
cross-spectral vectors can partially cancel, reducing the amplitude of the
total CS and hence the coherence.

A clear example is provided by Swift~J1727.8$-$1613
\citep[see Fig.~B.1 of][]{jin2026black}. When the Type-C and Type-B QPOs
are both present, the coherence peaks near the centroid frequencies of
the two QPOs and decreases in the region where their profiles overlap.
At the peak of the soft X-ray flare, when the Type-C QPO disappears and
the Type-B QPO dominates, the coherence instead shows a single, stronger
peak.

GX~339$-$4 shows a similar feature. In Fig.~\ref{coh}, the  coherence increases near the QPO centroid, shows a shallow decrease between the QPO and shoulder frequencies, and then increases again around the shoulder. This behavior is consistent with the frequency overlap of the two fitted components, which have different phase lags. By contrast, in observations where no shoulder component is required, the coherence remains close to unity across the QPO region.  The coherence spectrum therefore provides additional support for the presence of distinct contributions around the QPO and shoulder frequencies.

\subsection{Energy dependence of the phase lag and rms}

The energy-resolved fits provide an additional characterization of the
two fitted contributions (see Fig.~\ref{fig:lag_energy}). When the shoulder is resolved, the QPO lag relative to the 2.0$-$5.7~keV reference band remains small across the PCA band, usually below $\sim0.2$ rad. In several observations it is consistent with zero, or slightly negative, at the highest energies. The shoulder behaves differently: its lag is already larger than the QPO lag in the lowest non-reference band and generally increases with energy, reaching values of order $\sim0.8$$-$$1.7$ rad in the highest PCA band, where the uncertainties are largest.

The rms-energy spectra give a complementary result. The QPO fractional rms generally increases with photon energy and then flattens at higher energies, broadly consistent with that reported by \citet{zhang2017evolution} for GX~339$-$4 (see their Fig.~6). Similar rms$-$energy shapes have also been observed in other black-hole X-ray binaries, including Swift~J1727.8$-$1613 \citep{yu2024timing,zhu2024energy,yang2024timing,rawat2025evolution}, MAXI~J1803$-$298 \citep{zhu2023timing}, and MAXI~J1535$-$571 \citep{huang2018insight,garg2022energy,rawat2023comptonizing}. 

A rising rms--energy spectrum is commonly associated with variability of the Comptonized emission, with the lower-energy rms diluted by a less variable thermal disc component; the spectrum then flattens once the Comptonized component dominates \citep{sobolewska2006spectral,zycki2007modelling,gilfanov2009x}. Time-dependent Comptonization models can reproduce such rms and lag behavior through variations of the corona and its radiative coupling to the disc \citep{2022MNRAS.515.2099B,rawat2023comptonizing,alabarta2025geometry}.

For the same 2006/2007 rising phase of GX~339$-$4,
\citet{zhang2017evolution} found that the QPO lag--energy spectrum evolves from an
approximately monotonic increase below a QPO frequency of $\sim1.7$~Hz
to a more structured profile at higher QPO frequencies, including a broad
feature around the Fe-line energy that they associated with reflected
emission. Our component-resolved spectra show that the QPO fundamental retains structure around $\sim6-8$~keV in several observations, suggesting that part of the previously measured lag structure is intrinsic to the QPO. However, after separating the QPO and the shoulder, the strongest high-energy hard lag in the QPO region is carried by the shoulder (Figs.~\ref{fig:lag_energy} and \ref{fig:lag_energy_sep}).

This affects the interpretation of lag-energy spectra measured over a fixed QPO frequency interval. The traditional lag at the QPO frequency is obtained by averaging the CS over a finite interval around the QPO centroid. Once the shoulder is present, that interval can contain two nearby components with different phases, and the measured lag-energy spectrum is then the phase of the summed cross-vector, with a weight that depends on both Fourier frequency and photon energy. A hard lag measured over the QPO region therefore need not be the lag of the QPO fundamental alone. The comparison with \citet{zhang2017evolution} should accordingly be qualitative rather than one-to-one: their lags were measured relative to the 2$-$4~keV band and over the QPO region, whereas here the QPO fundamental and the shoulder are fitted as separate Lorentzian components in both the PDS and the CS.

The separation between the QPO and the shoulder is also consistent with recent component-resolved timing studies showing that the imaginary CS can reveal QPO-like components that are weak or blended in the PDS \citep{bellavita2025nature,fogantini2025hidden,jin2025timing,bollemeijer2025broad}. The GX~339$-$4 shoulder is not identical to the purely hidden cases, because it later becomes visible in the PDS.

A possible physical interpretation is that the QPO fundamental and the shoulder sample different parts of the spectral-timing response. The QPO lag structure around $\sim6$$-$$8$~keV could indicate a reflected-emission contribution, as suggested by \citet{zhang2017evolution}. The shoulder, by contrast, is more strongly delayed at high energies and may be more closely associated with the hard Comptonized response, through delayed feedback in an extended Comptonizing region \citep{karpouzas2020comptonizing,2022MNRAS.515.2099B} or inward propagation of accretion-rate fluctuations \citep{kotov2001x,2006MNRAS.367..801A}. The \textit{RXTE} data do not distinguish between these scenarios.

The highest-energy points should not be over-interpreted individually. Above $\sim20$~keV, the PCA count rate decreases and the energy-resolved CS becomes less well constrained, so the largest shoulder lags carry larger uncertainties than the lower-energy measurements. The shoulder already shows a harder lag than the QPO fundamental at lower energies, however, and the same separation appears in all observations where the shoulder is resolved.

\subsection{A possible Type-B-like precursor}
 
Interestingly, \cite{motta2009evolution} and \cite{gao2014type} reported a type-B QPO in the observation following the last observation in our sample, corresponding to the green data point just after the red shaded region in Figure~\ref{afig:lc}, based on the conventional PDS-based classification.
This QPO had a centroid frequency of $\sim $6.7 Hz and a fractional rms amplitude of $\sim $ 4.96\%. 
Its phase lag and rms$-$energy spectra are shown in Appendix~\ref{app:Energy} and Fig.~\ref{fig:lag_energy_typeb}. 
After the shoulder becomes detectable, its lag$-$energy behavior is broadly consistent with that of this Type-B QPO in nearly all observations: the phase lag generally increases with photon energy, although in some cases, such as Obs 21, the increase tends to flatten at higher energies. 

The shoulder also shows a clear temporal evolution in its rms$-$energy behavior. 
When it first becomes resolved, its fractional rms is well below that of the Type-C QPO, especially at high energies. 
As the source evolves, the shoulder rms gradually approaches the QPO rms in the high-energy bands; by Obs 20 the two components have comparable amplitudes, and in Obs 21 the shoulder becomes slightly stronger than the QPO over part of the energy range. 
This observation also corresponds to one of the lowest QPO phase lags in our sample (see the left panel of Fig.~\ref{fig:lag_energy}). 
In the final observation, the shoulder weakens again relative to the QPO. 
Despite these changes in relative amplitude, the shoulder generally follows an rms$-$energy dependence similar to that of the QPO, with the fractional rms increasing toward higher energies. 
Its rms$-$energy shape is less regular in the earliest stages, but becomes progressively better defined and gradually resembles the Type-B pattern shown in the right panel of Fig.~\ref{fig:lag_energy_typeb}.

In the conventional classification, Type-B QPOs are typically observed in the SIMS and are associated with relatively weak red-noise variability, total fractional rms amplitudes of about $5$--$10$\%, and $Q\geq6$ \citep{belloni2005evolution,casella2005abc,motta2011low,belloni2016transient}. Recent cross-spectral studies, however, suggest that their emergence may be more gradual. In Swift~J1727.8$-$1613, \citet{jin2026black} identified a broad component on the high-frequency side of the Type-C QPO during the HIMS that subsequently narrowed and became the dominant Type-B QPO in the SIMS. Its distinct phase lag and Type-B-like lag$-$energy spectrum suggest a possible evolutionary connection between the HIMS shoulder and the later Type-B QPO.
Type-C and Type-B QPOs have also been observed simultaneously over short intervals during state transitions \citep{motta2012discovery,motta2014black,rout2023x,wang2026transitional,jin2026black}. This indicates that the two modes need not replace one another instantaneously and may briefly coexist as the source changes state.

In GX 339$-$4 RXTE observations, the shoulder does not always satisfy the canonical narrow-QPO criterion expected for Type-B QPOs in the SIMS. 
In particular, it is broader than the subsequent Type-B QPO, with a quality factor of approximately $Q\sim3$ in most of the observations where the shoulder is detected (see Fig.~\ref{q}). 
It also appears on the high-frequency side of the Type-C QPO, rather than as an isolated narrow feature in the PDS. 
Thus, its PDS morphology alone would not justify a direct classification as a canonical Type-B QPO. 
Nevertheless, the similarity of the shoulder lag--energy and rms--energy
spectra to those of the subsequent Type-B QPO suggests that the two may
be related. Indeed, these spectral-timing behaviors are highly reminiscent of the Type-B QPO properties in GX~339$-$4 detailed by \citet{stevens2016phase}, who reported substantial hard lags and strong spectral modulations through phase-resolved spectroscopy. Furthermore, these energy-dependent profiles are qualitatively compatible with the
dual-corona Comptonization framework proposed by \citet{garcia2021two} and \citet{peirano2023dual}, which explains the Type-B lag and rms behavior via coherent oscillations within an extended coronal structure.
These similarities suggest that the shoulder may be
related to the Type-B QPO observed later in the outburst. The shoulder could
represent a broader, less developed form of the same variability,
already present during the late HIMS and becoming more prominent as the
source approached the Type-B-QPO interval. 

\section{Conclusions}
\label{conclusion}

We revisited the RXTE observations of GX~339$-$4 during the
rising phase of its 2006/2007 outburst and applied a joint PDS--CS
multi-Lorentzian decomposition to the type-C QPO region. Within this
framework, the QPO region is described by a narrow type-C QPO and a
neighboring high-frequency shoulder. Our main results are as follows.

\begin{enumerate}

\item The shoulder first appears at MJD~54142.04, when it is detected
mainly through the imaginary part of the CS and the local phase-lag
structure. It later becomes resolved in the PDS.

\item The shoulder follows the QPO frequency evolution, with
$\nu_{\rm sh}/\nu_{\rm QPO}\simeq1.04$--$1.18$, but carries a much larger
hard lag: $\sim0.5$--$0.8$~rad compared with $\lesssim0.17$~rad for the
QPO. Frequency drift may affect the QPO profile, but does not fully
describe the corresponding cross-spectral behavior. 

\item The energy-resolved fits show the same contrast. The QPO lag
remains small across most of the PCA band, whereas the shoulder lag is
larger and generally increases with energy. Their rms--energy spectra
are broadly similar, although their relative strengths evolve with time.

\item The physical origin of the shoulder remains uncertain. Its timing
properties resemble those of the type-B QPO detected later in the
outburst, suggesting that it may represent an earlier and broader stage
of the same variability. 

\item  In observations with sufficiently high signal-to-noise ratios, the
coherence is high at both the QPO and the shoulder frequencies but decreases
between them. This behavior is consistent with decomposing the QPO
feature into two components with different phase lags and differs from
the approximately constant unit coherence predicted for a common driving
signal acting through two deterministic transfer functions.
\end{enumerate}

\begin{acknowledgements}
We sincerely thank the anonymous referee for the careful reading of the manuscript and for the constructive comments, which helped improve its clarity and quality.
MM acknowledges the research
programme Athena with project number 184.034.002, which is (partly)
financed by the Dutch Research Council (NWO). 
Haifan~Zhu acknowledges support from the China Scholarship Council (CSC; Grant No.~202506270166). 
Wei Wang thanks the NSFC (12133007) and the National Key Research and Development Program of China (Grants No. 2021YFA0718503 and 2023YFA1607901) for support.
\end{acknowledgements}

\bibliographystyle{aa}
\bibliography{allref}

\begin{appendix}
\section{Light Curve of GX~339$-$4}
\setcounter{figure}{0}
\label{app:LIGHT}
To place our observations in the context of the 2006/2007 outburst evolution of GX~339$-$4, we examined the RXTE count rate, the hardness ratio and the fractional rms amplitude. As shown in Figure~\ref{afig:lc}, the observations selected for our analysis, marked by the red shaded region, occur during the rising phase of the outburst. Over this interval, the source brightens while both the hardness ratio and fractional rms amplitude decrease. The data is taken from \citet{motta2009evolution}; a more detailed discussion of the evolution of the outburst can be found there.

\begin{figure}[htbp]
\centering
    \includegraphics[scale=0.4]{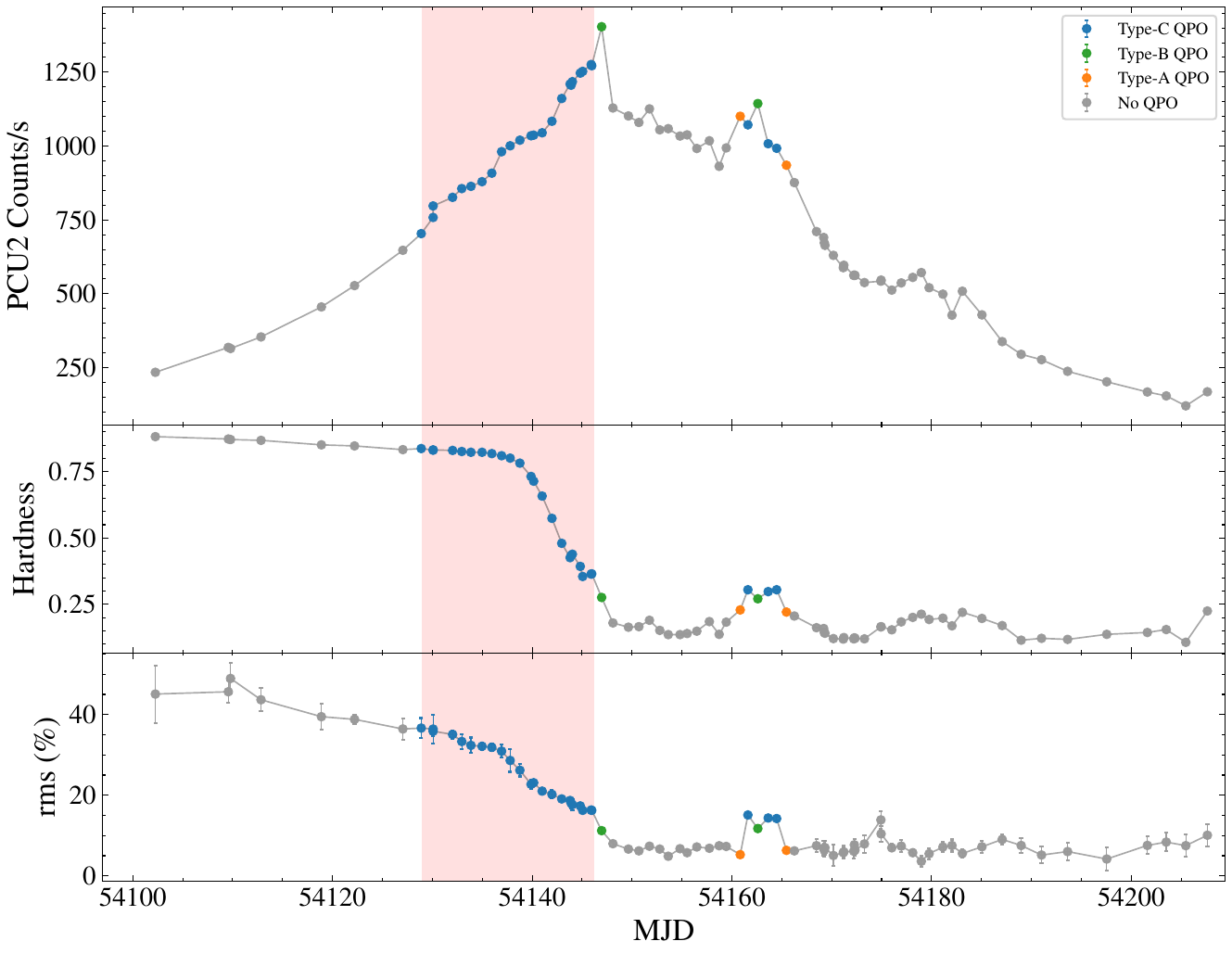}
\caption{
RXTE Light curve of GX 339$-$4 during the 2006/2007 outburst. The panels show, from top to bottom, the count rate, hardness ratio, and fractional rms amplitude. Blue, green, orange, and grey points indicate observations with type-C, type-B, type-A, and no detected QPO, respectively. The red shaded region marks the observations selected for our analysis.
}
\label{afig:lc}
\end{figure}
\section{Supplementary joint PDS$-$CS fits}
\label{app:fits}
This appendix presents supplementary fitting products for the representative observations discussed in Section~\ref{Results}. These figures show the soft- and hard-band PDS decompositions used in the joint PDS$-$CS analysis, together with the corresponding residuals and, where relevant, the intrinsic coherence spectra derived from the best-fitting models. They are intended to complement the main text figures, where we focus on the full-band PDS, the real and imaginary parts of the CS, and the derived phase lag spectra.

\begin{figure}[htbp]
\centering
    \includegraphics[scale=0.5]{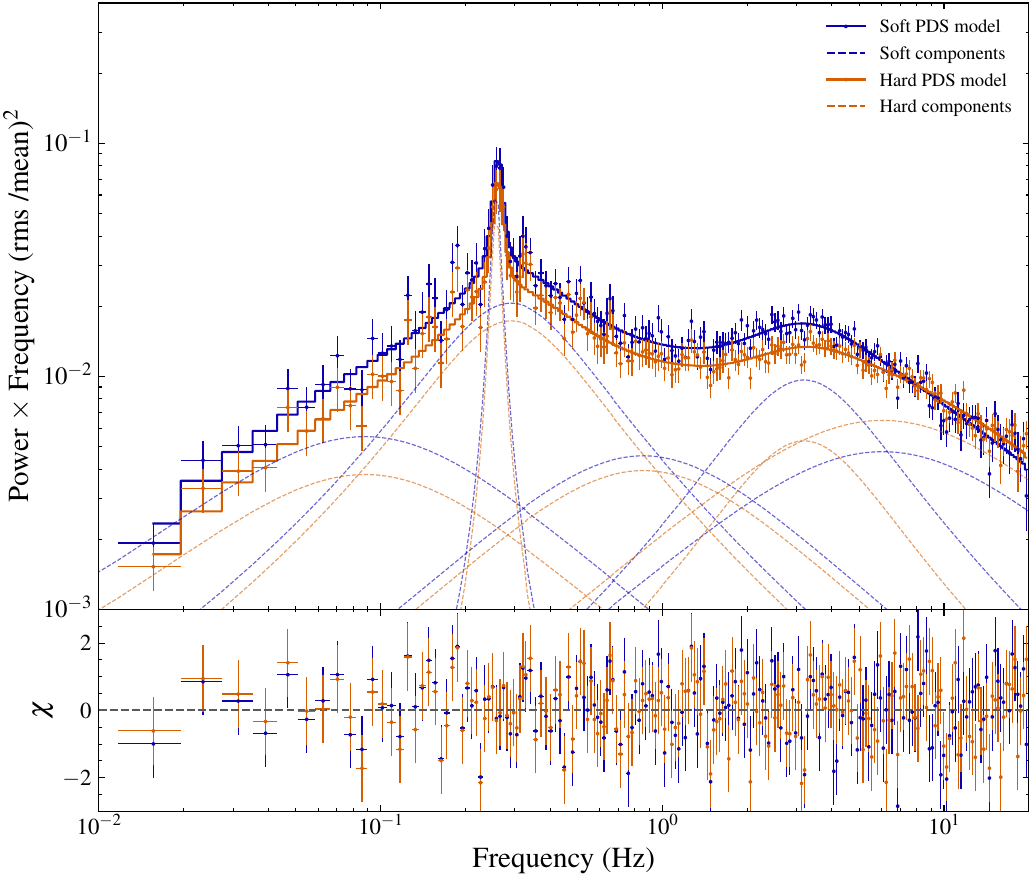}
\caption{
Supplementary PDS decomposition for Obs 6 of GX 339$-$4  at MJD~54133.92. 
The upper panel shows the soft- and hard-band PDS, plotted as power $\times$ frequency, together with the best-fitting multi-Lorentzian models. 
Solid curves show the total models, and dashed curves show the individual Lorentzian components. 
The lower panel shows the corresponding residuals. 
This observation is representative of the early rising phase, before the high-frequency shoulder is required.
}
\label{afig:fit1}
\end{figure}

\begin{figure}[htbp]
\centering
    \includegraphics[scale=0.4]{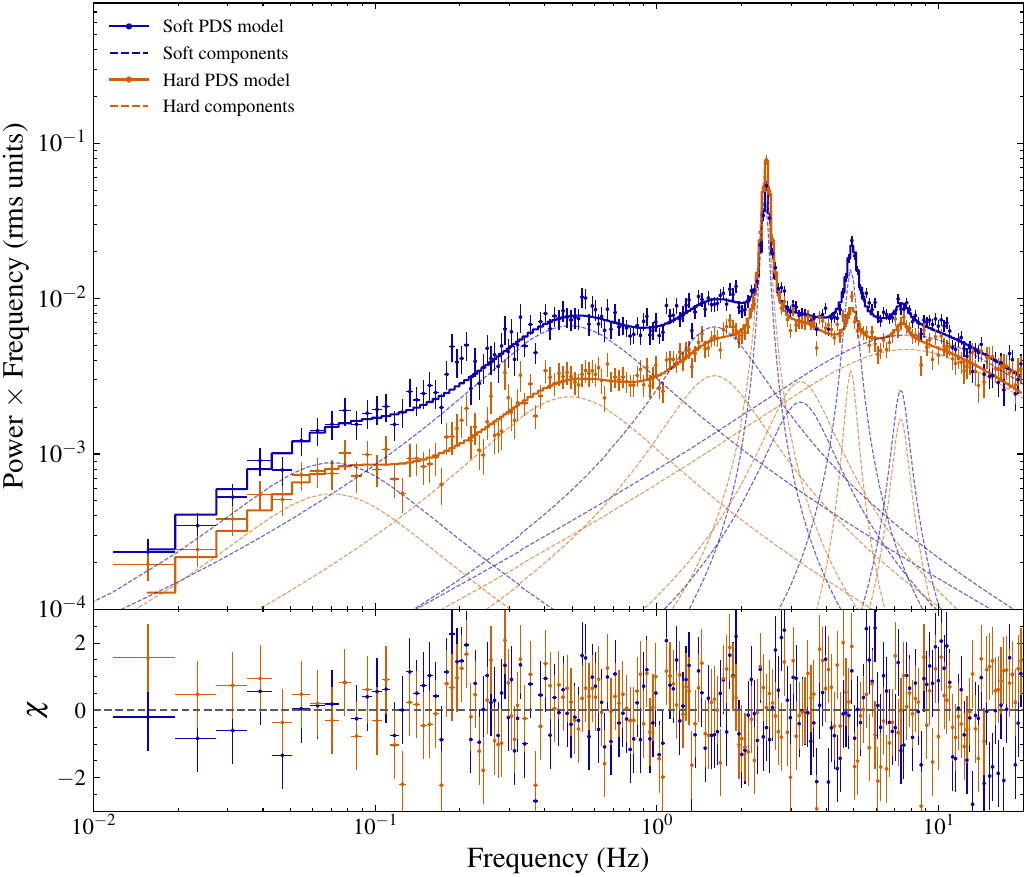}\\
    \includegraphics[scale=0.4]{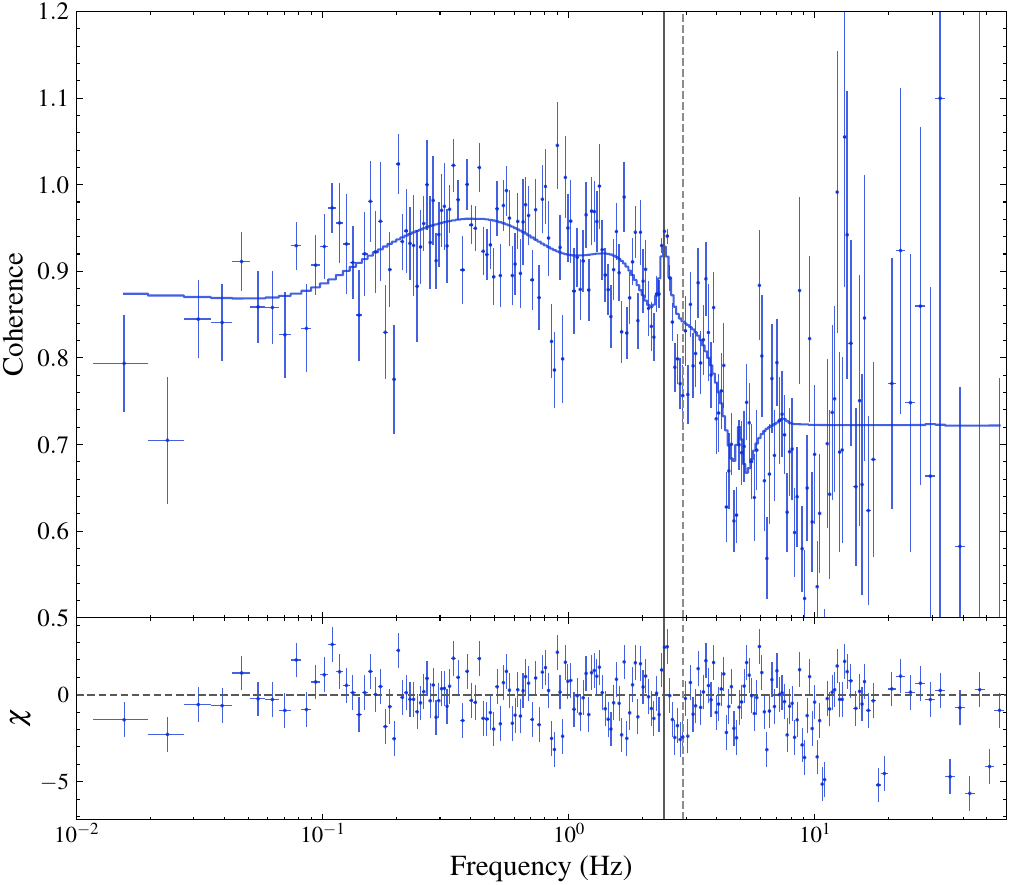}
\caption{
Same format as Fig.~\ref{afig:fit1}, but for Obs 15 of GX 339$-$4 at MJD~54142.04 using the 8-Lorentzian model. 
The lower panel shows the intrinsic coherence spectrum derived from the same joint PDS$-$CS model, together with the corresponding residuals. 
The vertical dashed lines mark the centroid frequencies of the QPO-related components. 
Although this model describes the PDS satisfactorily, it leaves structured residuals in the imaginary part of the CS and does not fully reproduce the local phase lag feature shown in Fig.~\ref{fig:qpofit3}. The coherence spectrum is displayed up to 60~Hz for completeness,
although only the 0.01--20~Hz range was included in the joint fit.
At the highest frequencies, the measurements are increasingly
dominated by statistical uncertainties.}

\label{afig:fit2}
\end{figure}

\begin{figure}[htbp]
\centering
    \includegraphics[scale=0.4]{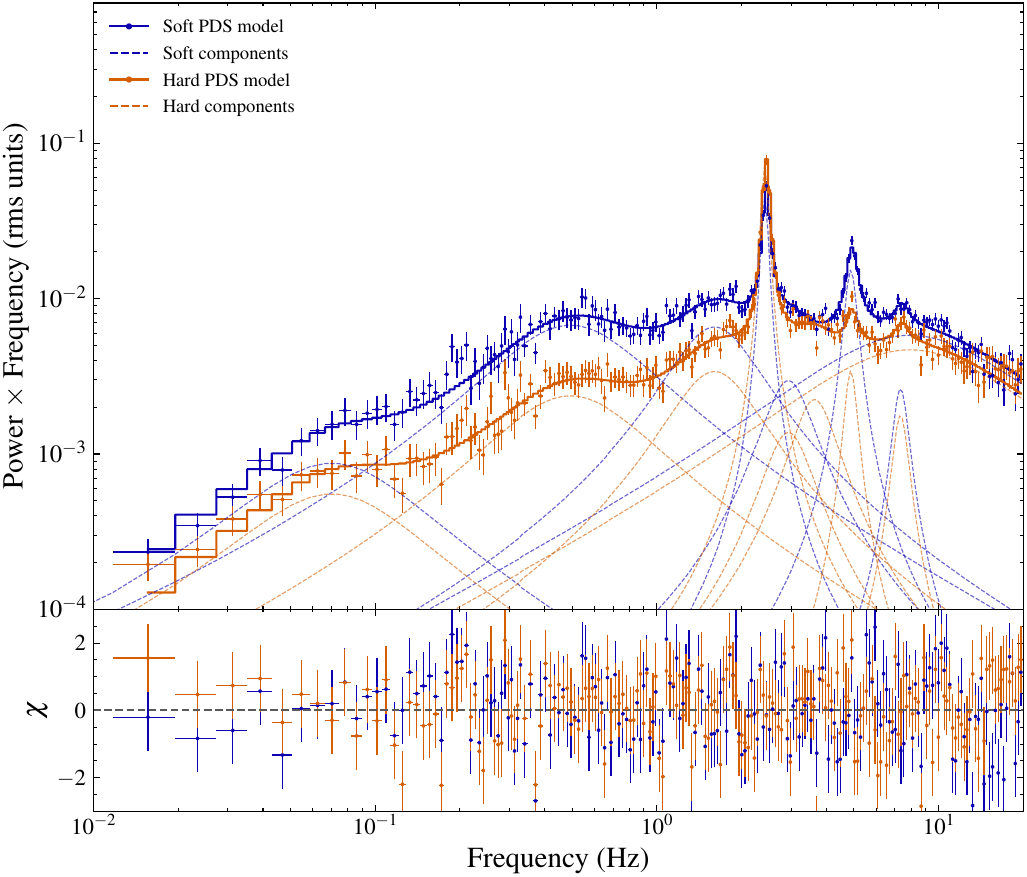}\\
    \includegraphics[scale=0.4]{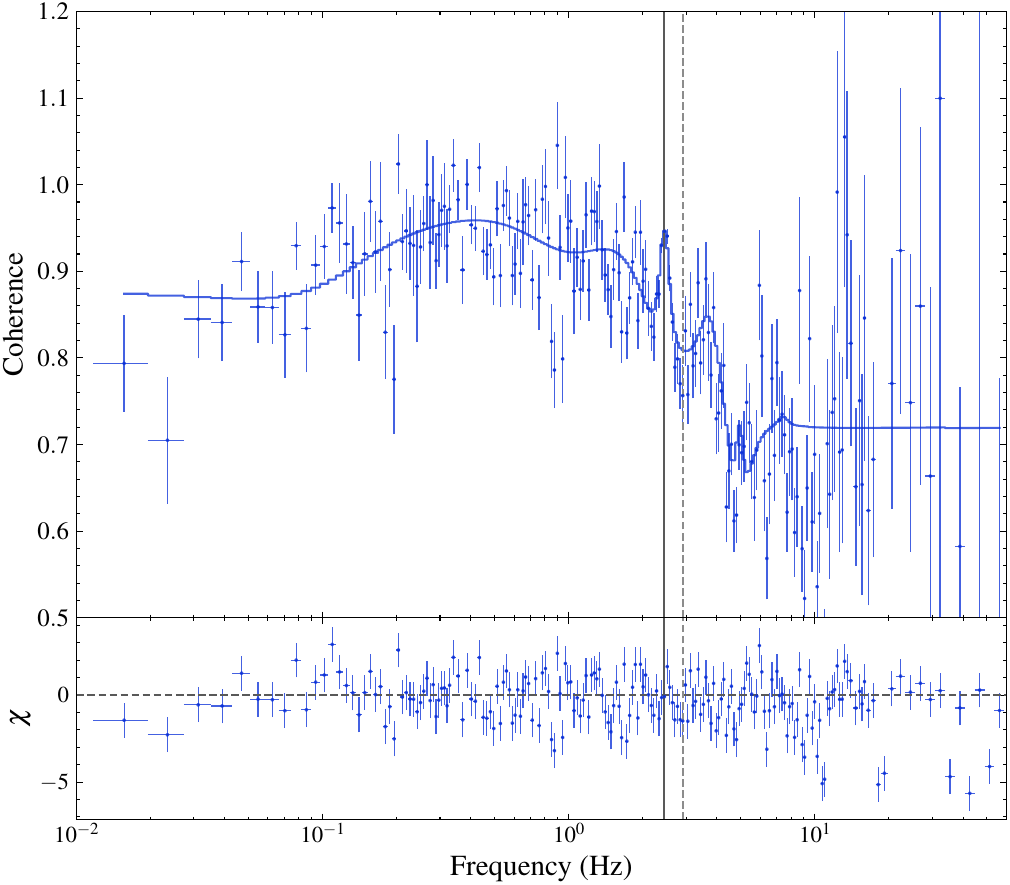}
\caption{
Same as Fig.~\ref{afig:fit2},but for the 9-Lorentzian model
including an additional shoulder-like Lorentzian component in the QPO
region. The added component improves the prediction of the coherence
function in the QPO--shoulder region, indicated by the dashed lines in
the lower panel, compared with Fig.~B.2, and gives a more consistent
description of the cross-spectral quantities shown in Fig.~\ref{fig:qpofit3}.
}
\label{afig:fit3}
\end{figure}

\begin{figure}[htbp]
\centering
    \includegraphics[scale=0.5]{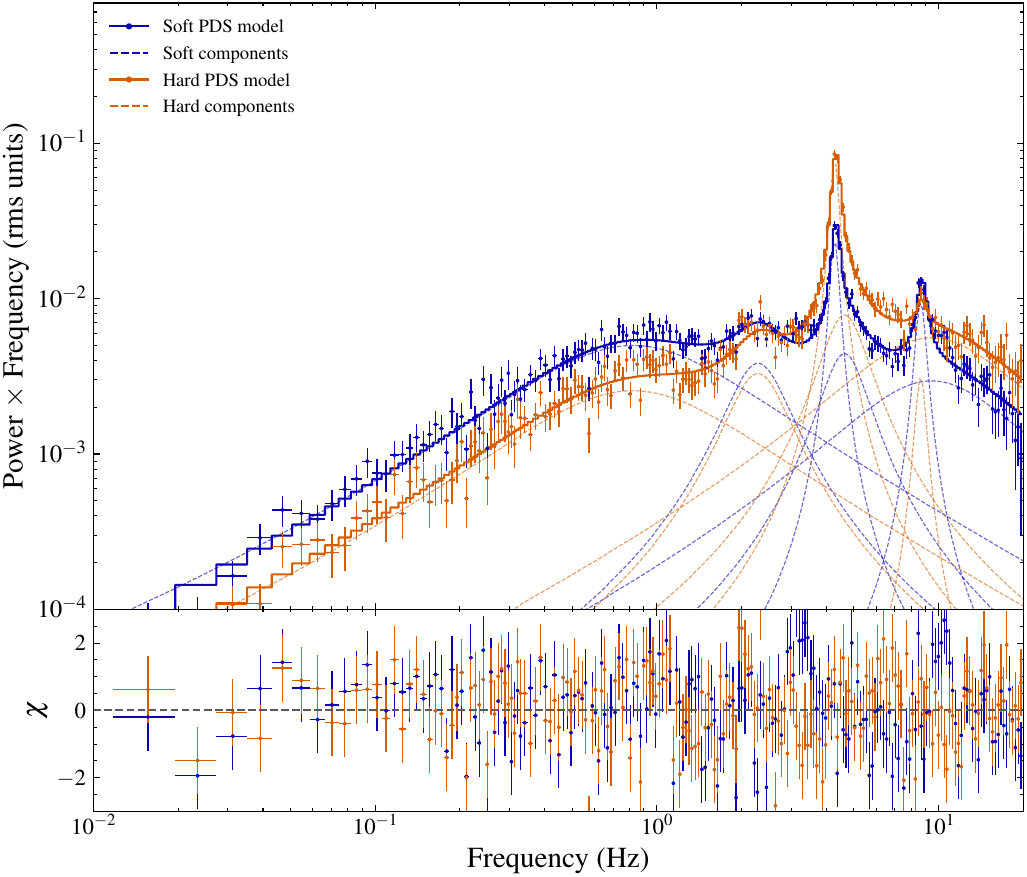}
\caption{
Same format as Fig.~\ref{afig:fit1}, but for Obs 17 of GX 339$-$4 at MJD~54143.87. 
In this observation, the high-frequency shoulder is clearly resolved on the high-frequency side of the QPO fundamental in both energy bands.
}
\label{afig:fit4}
\end{figure}

Fig.~\ref{afig:fit1} shows the early low-frequency observation used as a representative example before the shoulder is required. Figures~\ref{afig:fit2} and \ref{afig:fit3} show the transitional observation at MJD~54142.04, comparing the 8- and 9-Lorentzian descriptions. The 8-Lorentzian model already describes the soft- and hard-band PDS satisfactorily, while the 9-Lorentzian model includes the additional shoulder-like component required by the cross-spectral residuals discussed in Fig.~\ref{fig:qpofit3}. Fig.~\ref{afig:fit4} shows the following observation, where the shoulder is already visible as a resolved high-frequency component in the PDS. 

\begin{figure*}[htbp]
\centering
    \includegraphics[scale=0.35]{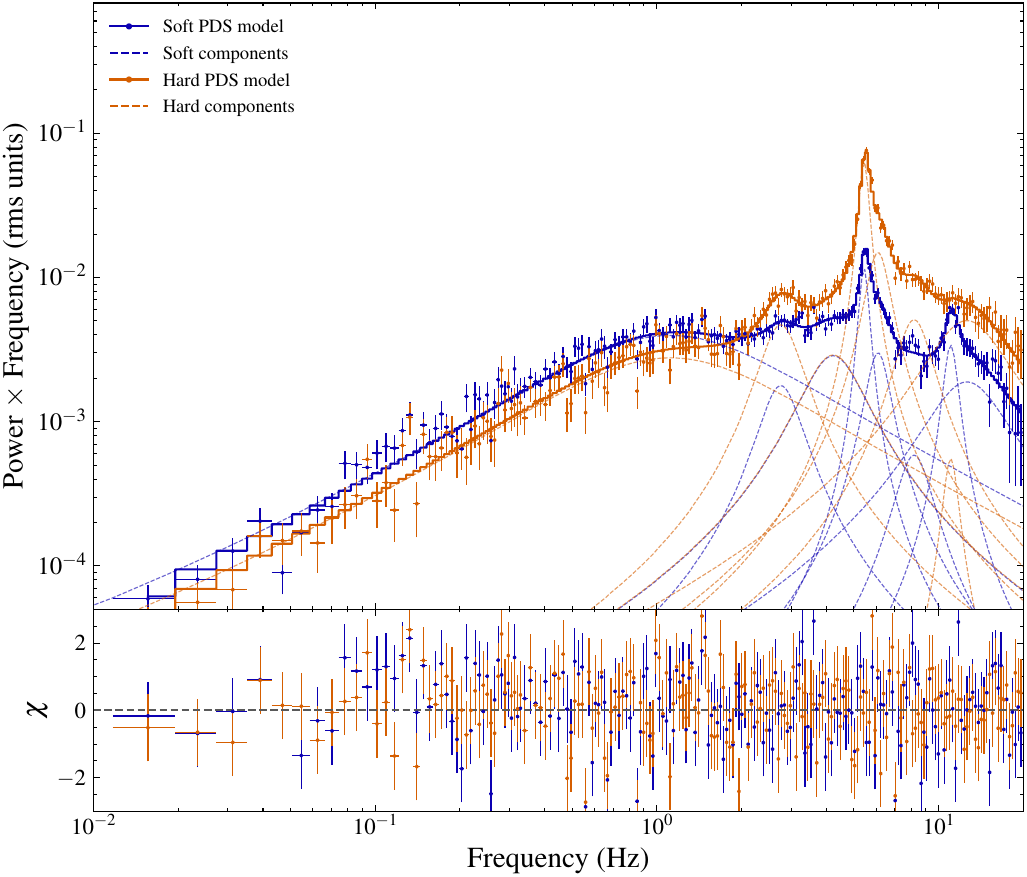}
    \includegraphics[scale=0.35]{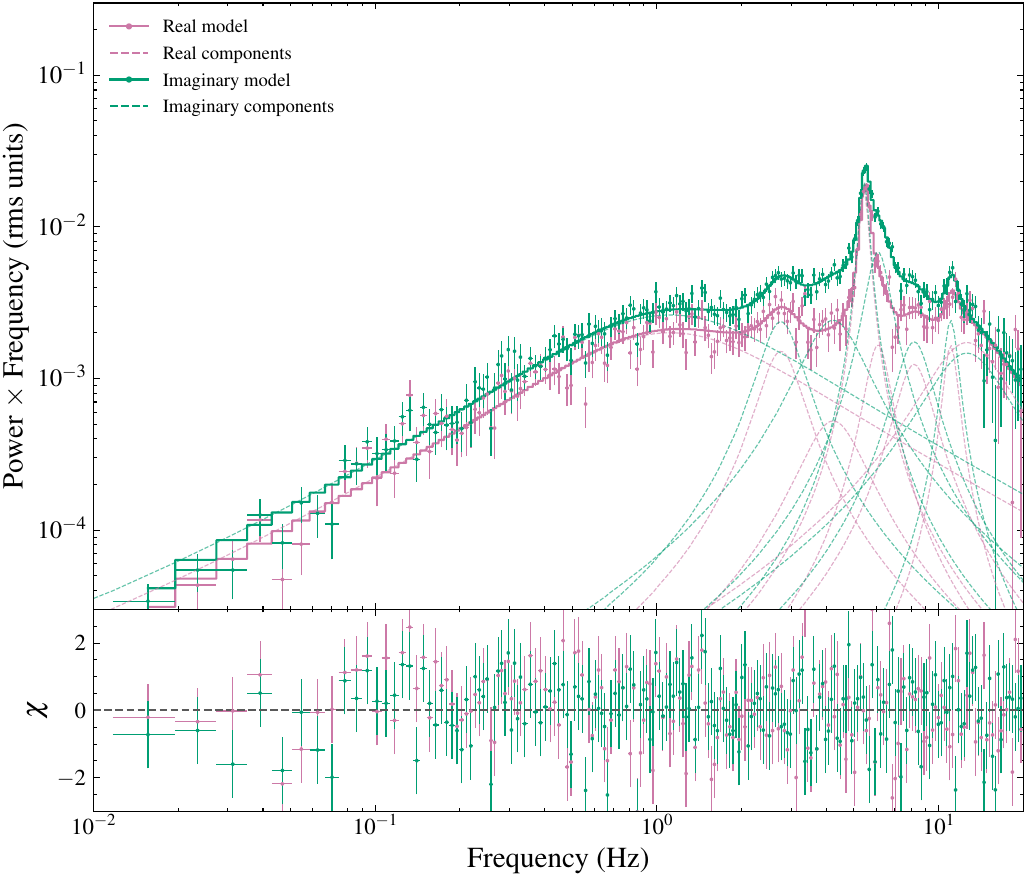}\\
    \includegraphics[scale=0.35]{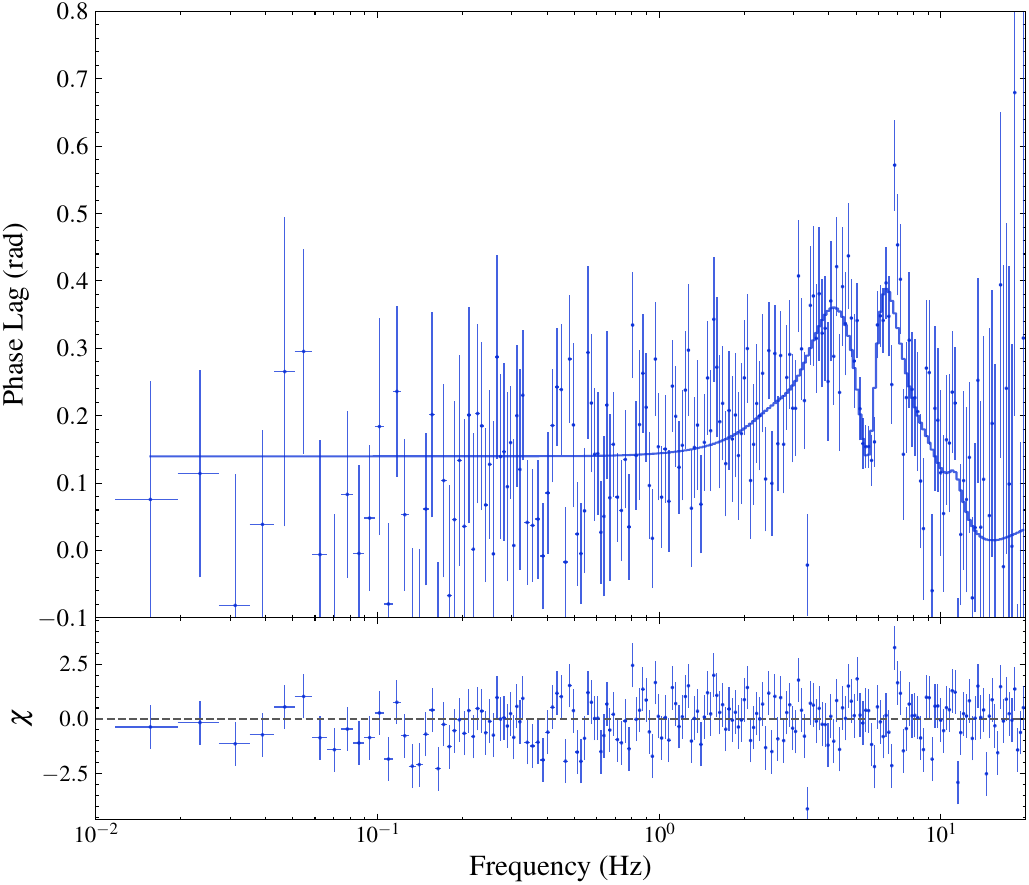}
    \includegraphics[scale=0.35]{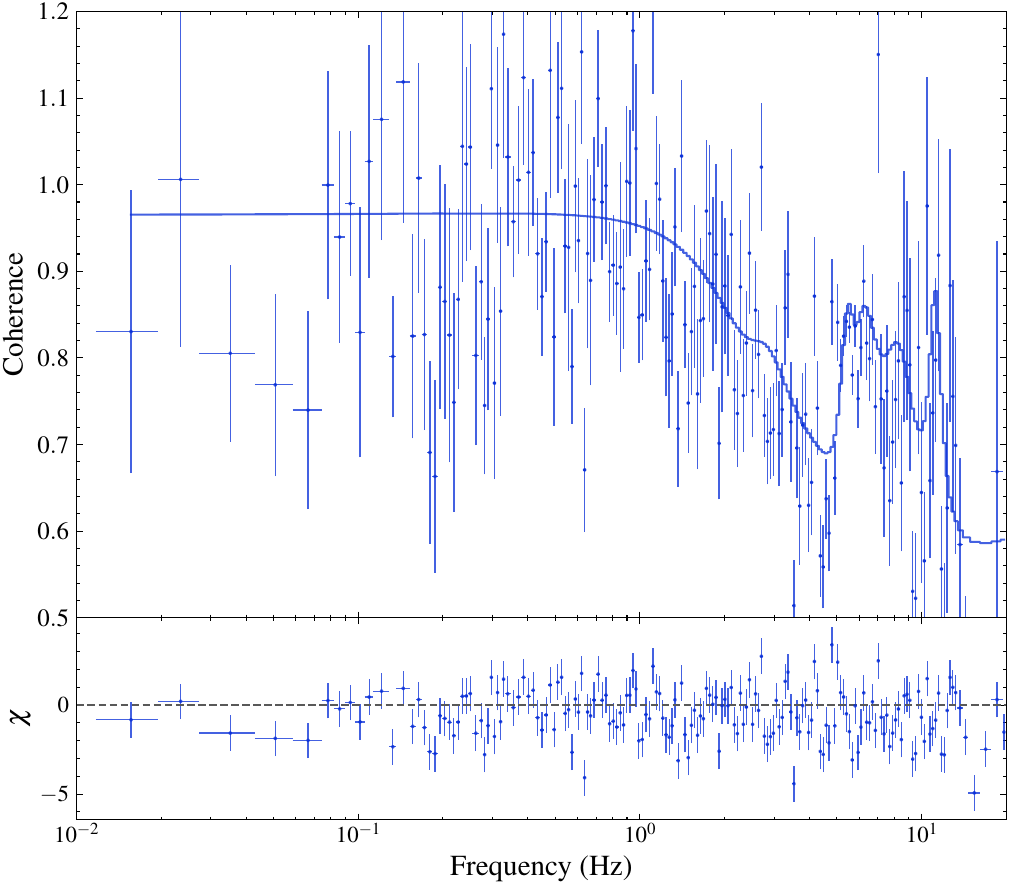}
\caption{
Same format as Fig.~\ref{afig:fit1}, but for Obs.~23 of GX~339$-$4 at MJD~54146.03. The QPO fundamental and the broader high-frequency shoulder are both resolved in the PDS, while the shoulder contributes more strongly to the imaginary part of the CS and carries a larger hard phase lag than the QPO. The fitted QPO and shoulder properties are broadly consistent with those reported by M24, as expected from the use of the same joint PDS--CS method. 
}
\label{afig:obs23}
\end{figure*}

\begin{figure*}[htbp]
\centering
    \includegraphics[width=0.7\textwidth]{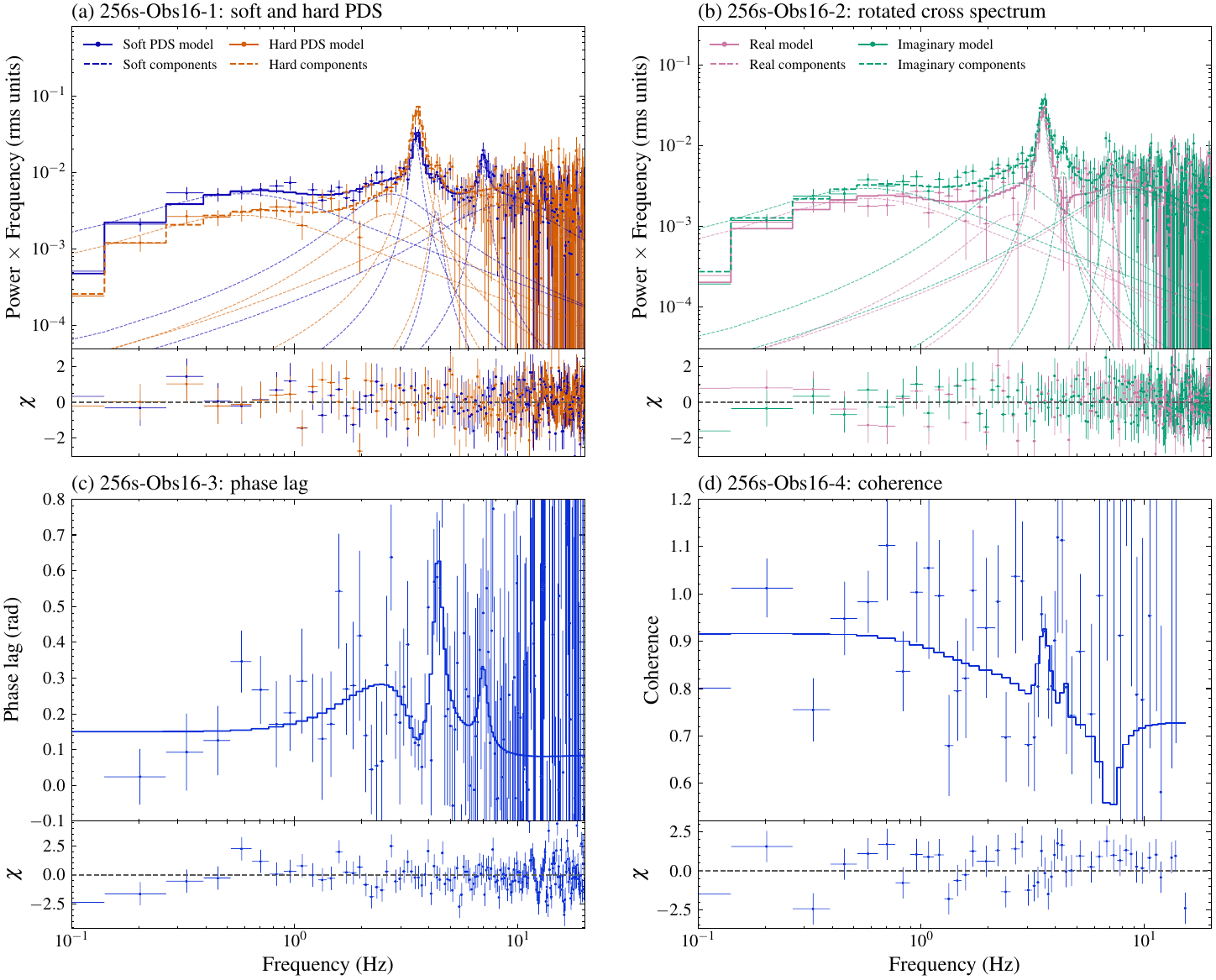}
\caption{Timing products of Obs.~16 calculated within a 256-s interval using
eight 32-s Fourier segments. Panels (a)--(d), from upper left to lower
right, show the soft- and hard-band PDS, the rotated real and imaginary
parts of the cross spectrum, the phase-lag spectrum, and the coherence,
respectively. Solid curves show the models, dashed curves in
panels (a) and (b) show the individual model components, and the lower
subpanels show the residuals.}

\label{app:short_cross}

\end{figure*}

Fig.~\ref{afig:obs23} presents the soft- and hard-band PDS, the real and imaginary parts of the CS, and the phase lag and coherence spectra for Obs.~23 of GX~339$-$4 at MJD~54146.03. 
Using the same data selection and energy bands as in
Section~\ref{obs}, we divided Obs.~16 into continuous 256-s
intervals, each containing eight 32-s Fourier segments. The soft- and
hard-band PDS and CS were averaged within each interval.
After subtracting the Poisson noise, the spectra were converted to
background-corrected fractional-rms normalization and rebinned to a frequency-bin width of 0.125 Hz by averaging four adjacent Fourier bins. We applied the same fixed
$45^{\circ}$ rotation as in the time-averaged analysis and fitted all timing products with the same multi-Lorentzian model. The resulting PDS and CS were fitted in XSPEC following the same procedure as the time-averaged analysis described in the main text. An example of the resulting fit is shown in Fig.~\ref{app:short_cross}.  The QPO harmonic is not significantly detected in the hard-band PDS, consistent with the results of \citet{axelsson2016revealing}, who found that it has a softer variability spectrum than the fundamental and becomes undetectable above $\sim10~\mathrm{keV}$ in the same observation; see their work for a detailed discussion of this energy dependence.

\begin{figure}[htbp]
\centering
    \includegraphics[width=0.4\textwidth]{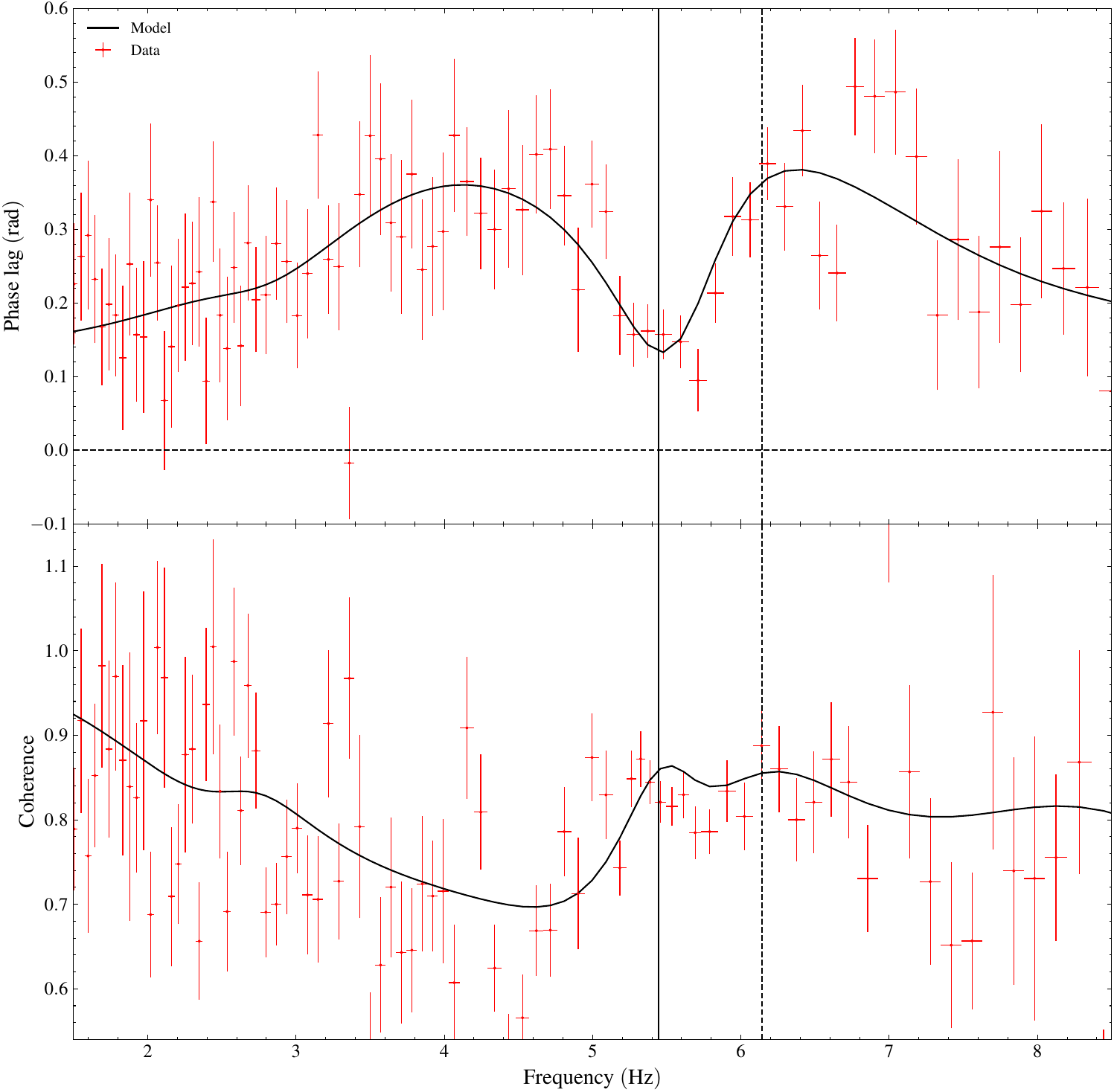}
\caption{
Phase lag spectrum (top) and coherence spectrum (bottom) of
GX~339$-$4 for Obs.~23 (ObsID~92035-01-03-06) over the QPO frequency
range. Red crosses show the measured values and the black curves show the profiles predicted
by the best-fitting joint PDS--CS model. The vertical solid and dashed
lines mark the centroid frequencies of the type-C QPO and the
high-frequency shoulder, respectively. The horizontal dashed line in the
upper panel marks zero phase lag. }

\label{coh}

\end{figure}

To better resolve the structure around the QPO frequency, we set the rebinning parameter to $-200$, thereby increasing the number of data points in this frequency range. The resulting spectra are shown in Fig.~\ref{coh}. For visualization, we applied additional moderate rebinning when plotting the data. The coherence spectrum clearly exhibits a double-peaked structure, with one peak at the QPO frequency and the other at the shoulder frequency.

\begin{figure}[htbp]
\centering
    \includegraphics[scale=0.4]{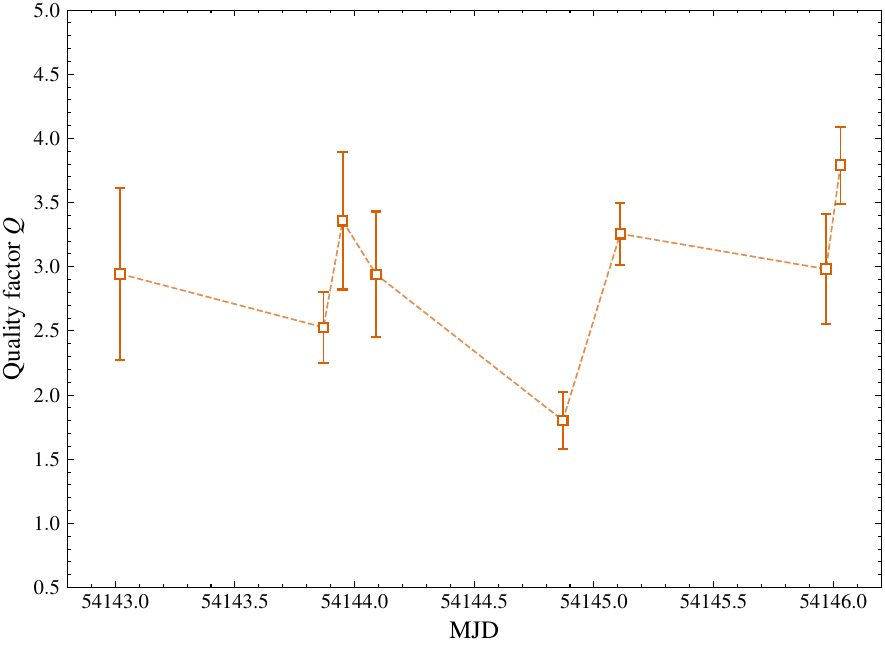}
\caption{
Evolution of the quality factor of the shoulder component during the rising phase of the 2006/2007 outburst of GX~339$-$4. 
The quality factor is defined as $Q=\nu_0/{\rm FWHM}$, where $\nu_0$ and ${\rm FWHM}$ are, respectively, the centroid frequency and full width at half maximum of the Lorentzian component obtained from the joint PDS$-$CS decomposition. 
The shoulder remains moderately broad, with $Q$ mostly in the range $\sim2.5$$-$$3.8$, except for one observation around MJD~54144.87 where $Q$ drops below 2. 
}

\label{q}

\end{figure}

Fig.~\ref{q} shows the temporal evolution of the quality factor of the shoulder component. 
The measured values are generally modest, mostly in the range $Q\sim2$$-$$4$, indicating that the shoulder remains relatively broad whenever it is resolved. 



\section{Energy dependence}
\label{app:Energy}
\begin{figure*}[htbp]
\centering
    \includegraphics[scale=0.45]{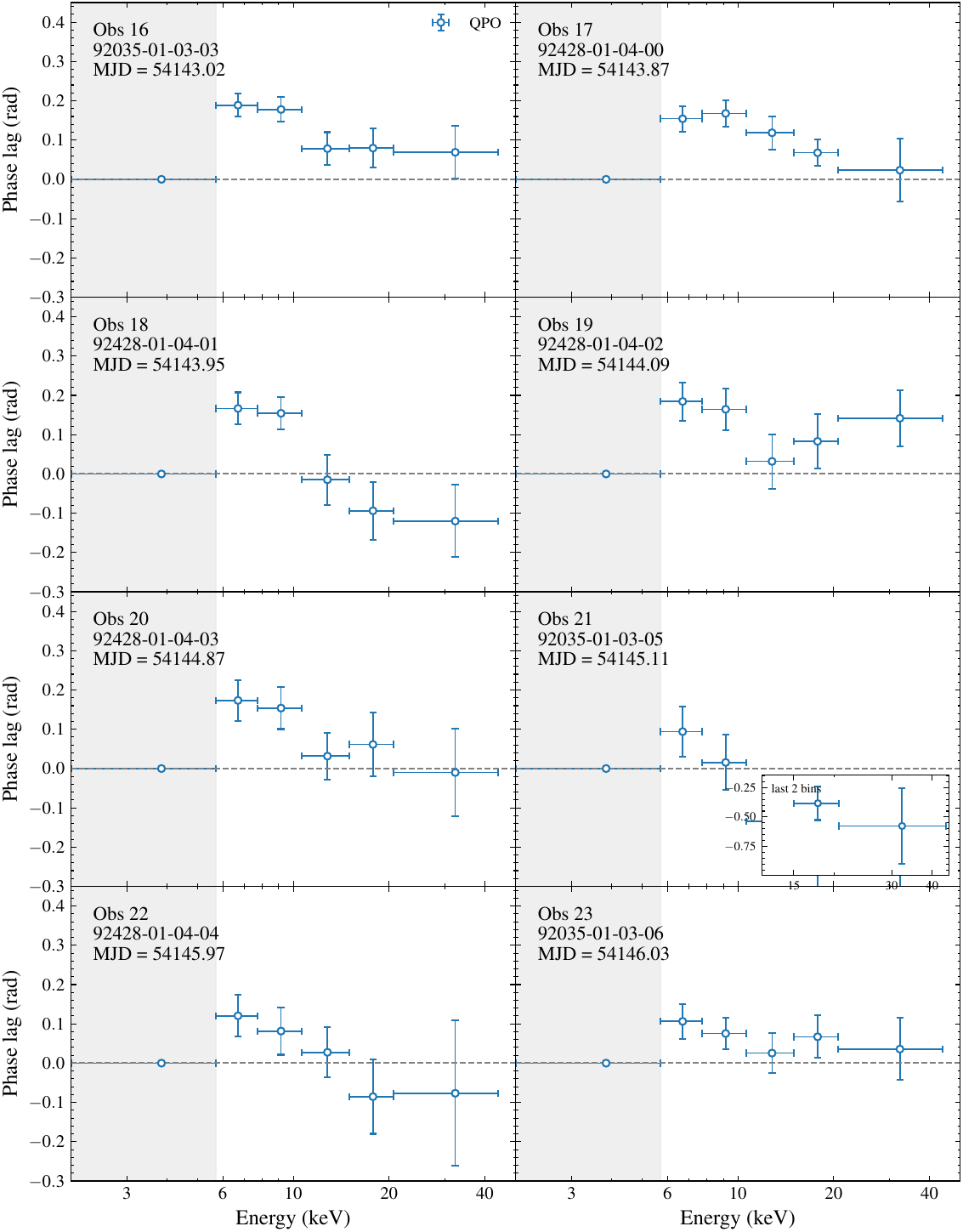}
    \includegraphics[scale=0.45]{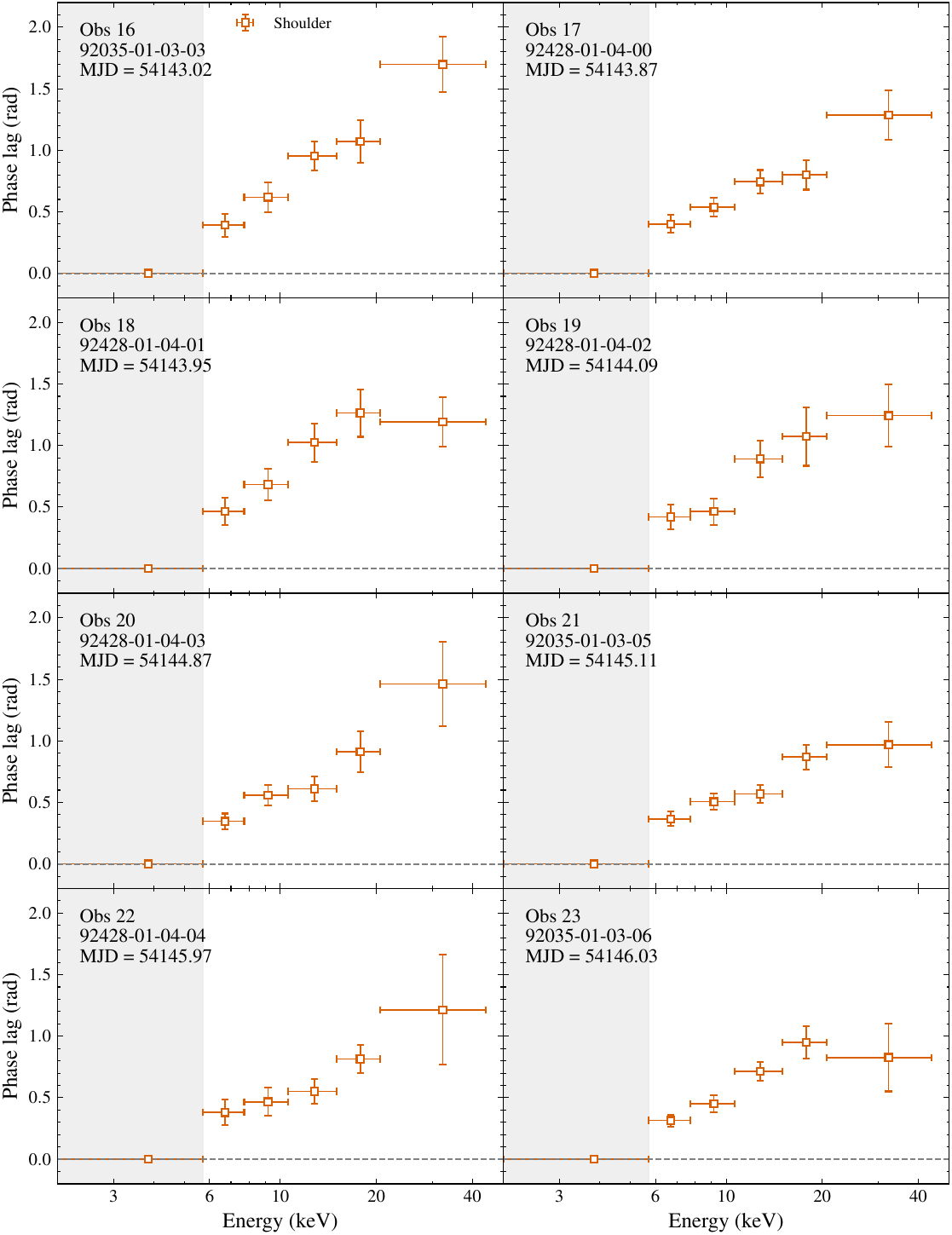}
\caption{
Energy dependence of the phase lags of the QPO fundamental and the QPO shoulder of GX 339$-$4. 
The left panel shows the lag-energy spectra of the QPO fundamental, while the right panel shows those of the shoulder component, for the observations in which the shoulder is clearly resolved. 
Each panel corresponds to one observation, with the observation number, ObsID, and MJD labeled. 
The grey shaded region marks the 2.0$-$5.7~keV reference band, for which the phase lag is fixed to zero by definition. 
}
\label{fig:lag_energy_sep}

\end{figure*}

Fig.~\ref{fig:lag_energy_sep} provides the separated lag-energy spectra of the QPO fundamental and the shoulder component. This figure is the same energy-resolved measurement shown in Fig.~\ref{fig:lag_energy}, but with the two components plotted separately to make their different vertical scales clearer. The QPO lags are confined to a narrow range around zero, so the left panel uses a smaller y-axis range. For Obs 21, the last two high-energy QPO points have larger uncertainties and extend below the main plotting range; these points are therefore shown in an inset. The shoulder lags are shown in the right panel with a wider y-axis range, because they are systematically larger and increase toward higher energies.

To place the energy-dependent behavior of the shoulder in context, we also examined the rms$-$energy and lag$-$energy spectra of a previously reported type-B QPO in GX~339$-$4. 
The measurements used here are taken from \cite{gao2014type}, who reported the fractional rms amplitudes and hard time lags in several energy bands for selected observations of GX~339$-$4. 
In particular, we use ObsID 92035-01-04-00 at MJD 54147.03, which was obtained shortly after the last observation (MJD 54146.03) in our sample and contains a type-B QPO at $\nu=6.70$ Hz.

\begin{figure*}[htbp]
\centering
    \includegraphics[scale=0.45]{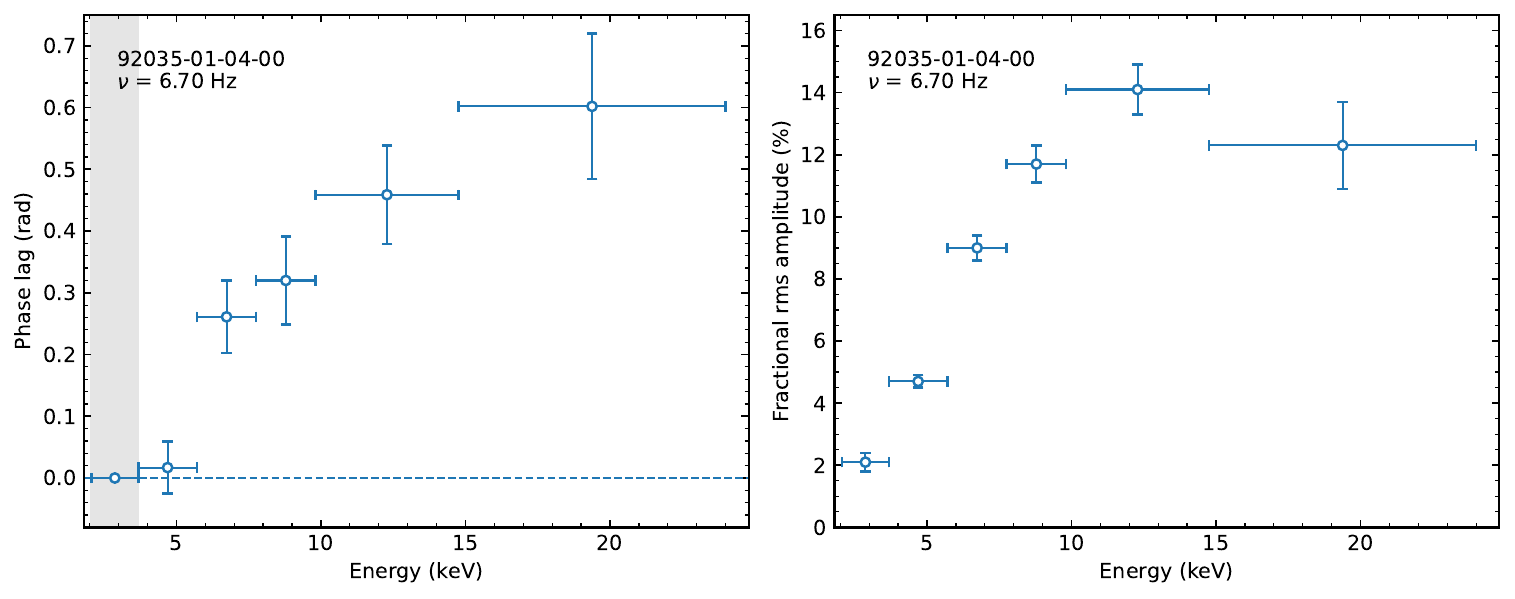}
\caption{
phase lag and rms$-$energy spectra for ObsID 92035-01-04-00 of the type-B QPO in GX 339$-$4 at MJD 54147.03. 
The time-lag and rms measurements are taken from \cite{gao2014type}. 
The phase lags are calculated from the reported time lags using $\phi = 2\pi\nu\tau$, with $\nu = 6.70$ Hz, and are measured relative to the lowest-energy reference band, 2.06$-$3.68 keV. 
The grey shaded region in the left panel marks the reference band, and the dashed line denotes zero lag. 
The right panel shows the fractional rms amplitude as a function of energy. 
}
\label{fig:lag_energy_typeb}

\end{figure*}

The rms amplitudes are directly taken from \cite{gao2014type}. 
The lag measurements in that work are given as hard time lags with respect to the lowest-energy band, 2.06$-$3.68 keV. 
For comparison with the phase lag quantities used in this paper, we converted the reported time lags into phase lags using
$    \phi = 2\pi \nu \tau $,
where $\nu=6.70$ Hz is the QPO centroid frequency and $\tau$ is the time lag in seconds. 
The resulting phase lag and rms$-$energy spectra are shown in Fig.~\ref{fig:lag_energy_typeb}.

The type-B QPO shows a clear increase of phase lag with photon energy. 
The rms amplitude also increases strongly from the soft band to intermediate energies, reaching a maximum of $\sim 14\%$ around 10$-$15 keV, before slightly decreasing in the highest-energy band. 
This behavior provides a useful reference for comparison with the shoulder component discussed in the main text, since both features show enhanced hard lags and increasing rms amplitude toward higher energies.

\end{appendix}

\end{document}